\documentclass{article}

\usepackage{tikz}
\usetikzlibrary{arrows}

\usepackage[ansinew]{inputenc}          %skandien syöttö (windows)
\usepackage{amssymb,amsthm,amsmath}     %ams
\usepackage{a4wide}

\theoremstyle{plain}
\newtheorem{theorem}{Theorem}%[section]

\newtheorem{corollary}[theorem]{Corollary}
\newtheorem{proposition}[theorem]{Proposition}
\theoremstyle{definition}
\newtheorem{definition}[theorem]{Definition}

\newcommand{\Z}{\mathbb{Z}}

\newcommand{\SG}{\mathcal{S}}
\newcommand{\TG}{\mathcal{T}}
\newcommand{\KG}{\mathcal{K}}

\newcommand{\ID}{\gamma^{ID}}
\newcommand{\LD}{\gamma^{LD}}
\newcommand{\SID}{\gamma^{SID}}

\begin{document}

%\title{Tolerant identifying collections of codes for sensor networks\footnote{The paper has been presented in part in the Bordeaux Graph Workshop 2016 (BGW 2016). An extended abstract~\cite{} of the paper has been published in the conference proceedings of the BGW 2016.}}

\title{Optimal Bounds on Codes for Location in Circulant Graphs}

\author{\textbf{Ville Junnila} and \textbf{Tero Laihonen} \\
Department of Mathematics and Statistics\\
University of Turku, FI-20014 Turku, Finland\\
viljun@utu.fi and terolai@utu.fi \and \textbf{Gabrielle Paris}\thanks{Supported by the ANR-14-CE25-0006 project of the French National Research Agency}\\
LIRIS, University of Lyon, France\\
%University of Turku, FI-20014 Turku, Finland\\
claudia.paris-sierra@univ-lyon1.fr}
\date{}
\maketitle

\begin{abstract} Identifying  and locating-dominating codes have been studied widely in circulant graphs of type $C_n(1,2,3,\dots, r)$  over the recent years. In 2013, Ghebleh and Niepel studied locating-dominating and identifying codes in the circulant graphs $C_n(1,d)$ for $d=3$ and proposed as an open question the case of $d > 3$. In this paper we study identifying, locating-dominating and self-identifying codes in the graphs $C_n(1,d)$, $C_n(1,d-1,d)$ and $C_n(1,d-1,d,d+1)$. We give a new method to study lower bounds for these three codes in the circulant graphs using suitable grids. Moreover, we show that these bounds are attained for infinitely many parameters $n$ and $d$. In addition, new approaches are provided which give the exact values for the optimal self-identifying codes in $C_n(1,3)$ and $C_n(1,4).$
\end{abstract}


%\noindent \textbf{Keywords:} Identifying code; location detection;
%sensor network; malfunctioning sensor; covering design

\section{Introduction}

Let $G = (V, E)$ be a simple, undirected graph with the vertex set $V$ and the edge set $E$. The \emph{open neighbourhood} $N(G;u)$ of $u \in V$ consists of the vertices adjacent to $u$, i.e., $N(G;u) = \{ v \in V \ | \ uv \in E \}$. The \emph{closed neighbourhood} $N[G;u]$ of $u \in V$ is defined as $N[G;u] = N(u) \cup \{u\}$. Regarding the open and closed neighbourhoods, if the underlying graph is known from the context, then we can simply write $N(G;u) = N(u)$ and $N[G;u] = N[u]$. A nonempty subset $C \subseteq V$ is called a \emph{code}, and its elements are called \emph{codewords}. The \emph{identifying set} (or the $I$\emph{-set} or the \emph{identifier}) of $u$ is defined as $I(G,C;u) = N[G;u] \cap C$; if the graph $G$ or the code $C$ is known from the context, then we can again write $I(G,C;u) = I(G;u) = I(C;u) = I(u)$.
The \emph{distance} between two vertices $u,v\in V$ is the number of edges in any shortest path between them and it is denoted by $d_G(u,v)=d(u,v)$.
A graph $G$ is $r$-\emph{regular} if $|N(G;u)|=r$ for all $u\in V$.

Let $C$ be a code in $G$. A vertex $u \in V$ is \emph{covered} or \emph{dominated} by a codeword of $C$ if the identifying set $I(C;u)$ is nonempty. The code $C$ is \emph{dominating} in $G$ if all the vertices of $V$ are covered by a codeword of $C$, i.e., $|I(C;u)| \geq 1$ for all $u \in V$. The code $C$ is \emph{identifying} in $G$ if $C$ is dominating and for all distinct $u,v \in V$ we have
%\begin{equation} \label{EQIDCondition}
%I(C;u) \neq I(C;v) \textrm{.}
%\end{equation}
\[
I(C;u) \neq I(C;v) \textrm{.}
\]
The definition of identifying codes is due to Karpovsky \emph{et al.}~\cite{kcl}, and the original motivation for studying such codes comes from fault diagnosis in multiprocessor systems. The concept of locating-dominating codes is closely related to the one of identifying codes. We say that the code is \emph{locating-dominating} in $G$ if $C$ is dominating and for all distinct $u,v \in V \setminus C$ we have $I(C;u) \neq I(C;v)$.
%\[
%I(C;u) \neq I(C;v) \textrm{.}
%\]
The definition of locating-dominating codes was introduced by Slater~\cite{RS:LDnumber,S:DomLocAcyclic,S:DomandRef}. The original motivation for locating-dominating codes was based on fire and intruder alarm systems. An identifying or locating-dominating code with the smallest cardinality in a given finite graph $G$ is called \emph{optimal}. The number of codewords in an optimal identifying and locating-dominating code in a finite graph $G$ is denoted by $\ID(G)$ and $\LD(G)$, respectively.

In this paper, we focus on studying identifying and locating-dominating codes (as well as self-identifying codes which are defined later) in so called circulant graphs. For the definition of circulant graphs, we first assume that $n$ and $d_1, d_2, \ldots, d_k$ are positive integers and $d_i\le n/2$ for all $i=1,\dots,k$. Then the circulant graph $C_n(d_1, d_2, \ldots, d_k)$ is defined as follows: the vertex set is $\Z_n = \{0,1, \ldots, n-1\}$ and the open neighbourhood of a vertex $u \in \Z_n$ is
\[
N(u) = \{u \pm d_1, u \pm d_2, \ldots, u \pm d_k\} \text{,}
\]
where the calculations are done modulo $n$. Previously, in~\cite{BCHLildchc,CLMcp,EJLldc,GMS:IdCyc,JLidcp,Mlldcn,RobRob,XTHicco}, identifying and locating-dominating codes have been studied in the circulant graphs $C_n(1,2, \ldots, r)$ $(r \in \Z, r \geq 1)$, which can also be viewed as power graphs of cycles of length $n$. Recently, in~\cite{GNlidcn}, Ghebleh and Niepel studied identification and location-domination in $C_n(1,3)$. They obtained the following results:
\[
\lceil 4n/11 \rceil \leq \ID(C_n(1,3)) \leq \lceil 4n/11 \rceil +1 \ \text{ and } \ \lceil n/3 \rceil \leq \LD(C_n(1,3)) \leq \lceil n/3 \rceil +1 \text{.}
\]
Moreover, they showed that in most cases the given lower bounds are actually the exact values of $\ID(C_n(1,3))$ and $\LD(C_n(1,3))$ and conjectured that in the rest of the cases the lower bound could be increased by one (attaining the given constructions). They also stated as an open question what happens in the graphs $C_n(1,d)$ with $d$ being greater than $3$ and mentioned that the methods used in their paper seem to be non-applicable. In this paper, we present a new approach to determine $\ID(C_n(1,d))$ and $\LD(C_n(1,d))$ with $d \ge 3$. The new approach is based on the observation that identification and locating-domination in the circulant graphs $C_n(1,d)$ have connections to identifying and locating-dominating codes in the infinite square grid. In particular, we can take advantage of the known lower bounds for identifying and locating-dominating codes in the square grid and derive lower bounds for the circulant graphs $C_n(1,d)$. Moreover, there exist similar connections and results between the circulant graphs $C_n(1,d-1,d)$ and $C_n(1,d-1,d,d+1)$ and the infinite triangular grid and king grid, respectively. In Section~\ref{SecGridsCirculants}, these connections as well as the needed definitions and known results regarding the grids are discussed, and we also present the lower bounds for the circulants graphs obtained from the grids. Then, in Section~\ref{SecIDLD}, we present constructions of identifying and locating-dominating codes for the circulant graphs. In particular, we obtain infinite families of circulant graphs with optimal identifying codes as well as families with optimal locating-dominating codes.

In addition to considering identification and location-domination, we also study  self-identifying codes, which overcome some issues of the regular identifying codes described in the following. Indeed, if $C$ is an identifying code in a graph $G = (V,E)$, then we can locate one irregularity (for example, a fire or an intruder) in $G$ as all the identifying sets are distinct. However, if there are more than one irregularity in $G$, then we can mislocate the irregularity (see \cite{+koodithl}), since we could have $I(C;u) = I(C;v_1) \cup I(C;v_2)$ for some vertices $u,v_1,v_2 \in V$, and more disturbingly not even notice that something is wrong. Thus, to locate one irregularity and detect multiple ones, the following definition of self-identifying codes have been introduced in~\cite{+koodithl} (although in the paper the code is called $1^+$-identifying). %Thus, to locate one irregularity and detect multiple ones, the following definition of self-identifying codes as well as more detailed motivation have been given in~\cite{+koodithl} (although in the paper the code is called $1^+$-identifying).
\begin{definition}
A code $C \subseteq V$ is \emph{self-identifying} in $G =(V,E)$ if for all distinct $u,v \in V$ we have
\[
I(C;u) \setminus I(C;v) \neq \emptyset \text{.}
\]
In a finite graph $G$, a self-identifying code with the smallest cardinality is called optimal, and the number of codewords in an optimal self-identifying code in $G$ is denoted by $\SID(G)$.
\end{definition}
The self-identifying codes have also been discussed in~\cite{JLcctl,JLtldsn}. In those papers, it has been shown that $C$ is a self-identifying code in $G$ if and only if for all $u \in V$ we have $I(C;u) \neq \emptyset$ and
\begin{equation} \label{CharacterizationSID}
\bigcap_{c \in I(C;u)} N[c] = \{u\} \text{.}
\end{equation}
Therefore, the sought vertex can be determined only using its identifying set; compare this to regular identifying codes where the identifying set has to be compared to other identifying sets in order to locate a vertex. In Sections~\ref{SecGridsCirculants} and \ref{SecSID}, we present results for self-identifying codes in the circulant graphs; especially, we focus on results in the graphs $C_n(1,d)$, $C_n(1,d-1,d)$ and $C_n(1,d-1,d,d+1)$.

\section{Infinite grids and circulant graphs} \label{SecGridsCirculants}

In this section, we first recall some preliminary definitions and results regarding infinite square, triangular and king grids and then present the connections between circulant graphs and grids. Let us first present the definitions of the grids. In all the grids, the vertex set is $V = \Z^2$. The edges of the \emph{square grid} $\SG$ are defined in such a way that the closed neighbourhood of $u = (x,y) \in \Z^2$ is
\[
N[\SG;u] = \{ (x',y') \in \Z^2 \ | \ |x-x'| + |y-y'| \leq 1 \} \text{.}
\]
The edges of the \emph{triangular grid} $\TG$ is defined in such a way that the closed neighbourhood of $u = (x,y) \in \Z^2$ is $N[\TG;u] = N[\SG;u] \cup \{(x+1,y+1), (x-1,y-1)\}$. The edges of the \emph{king grid} $\KG$ is defined in such a way that the closed neighbourhood of $u = (x,y) \in \Z^2$ is $N[\KG;u] = N[\TG;u] \cup \{(x-1,y+1), (x+1,y-1)\}$. For comparing the sizes of codes, we need a way to measure them in the infinite grids. For this purpose, we first denote
\[
Q_m = \{(x,y) \in \Z^2 \ | \ |x| \leq m, |y| \leq m \} \text{,}
\]
where $m$ is a positive integer. The \emph{density} of a code $C \subseteq \Z^2$ is then defined as
\[
D(C) = \limsup_{m \rightarrow \infty} \frac{|C \cap Q_m|}{|Q_m|} \text{.}
\]
For a finite nonempty set $S\subseteq V$ in a graph $G=(V,E)$, the (local) \emph{density} of a code $C\subseteq V$ in $S$ is defined as $|S\cap C|/|S|.$

Analogously to finite graphs, an identifying, locating-dominating and self-identifying code with the smallest density in the square, triangular or king grid is called \emph{optimal}. The densities of optimal codes on these grids have been intensively studied and all the exact values are known. The optimal densities can be found in Table~\ref{TableOptimalDensities} together with the references to the papers, where the results have been presented.

\begin{table}
\centering
\begin{tabular}{c|ccc}
  % after \\: \hline or \cline{col1-col2} \cline{col3-col4} ...
    & square grid $\SG$ & triangular grid $\TG$ & king grid $\KG$ \\
    \hline
    LD & $3/10$ \cite{S:fault-tolerant}& $13/57$ \cite{H:OptLDsetTri}& $1/5$ \cite{HLl-d}\\
    ID & $7/20$ \cite{BHL:ExtMinDenSqr,monta}& $1/4$ \cite{kcl}& $2/9$ \cite{chl,chlz2}\\
    self-ID & $1/2$ \cite{+koodithl}& $1/2$ \cite{+koodithl}& $1/3$ \cite{+koodithl}
\end{tabular}
\caption{The densities of optimal identifying (ID), locating-dominating (LD) and self-identifying (self-ID) codes in the square $\SG$, triangular $\TG$ and king grids $\KG$ are listed in the table. Next to each density you can find the reference to the result.} \label{TableOptimalDensities}
\end{table}

In the following theorem, we present the connection between  identifying, locating-dominating and self-identifying codes in the square grid and the circulant graphs $C_n(1,d)$.
\begin{theorem} \label{reduction}
Let $n$, $d$ and $k$ be positive integers such that $d\ge 2$. If $C$ is an identifying code in $C_n(1,d)$ with $k$ codewords, then there exists an identifying code in the infinite square grid $\SG$ with density $k/n$. Analogous results also hold for locating-dominating and self-identifying codes.
\end{theorem}
\begin{proof}
Let $G=C_n(1,d)$ be a circulant graph and $C$ an identifying code in it. We will use the following correspondence of the vertex $x=(x_1,x_2)\in \Z^2$ in the square grid with the vertex $x_1+ x_2\cdot d$ in $C_n(1,d)$ where  $x_1+x_2\cdot d$ is calculated modulo $n$ (throughout the paper). Namely, the closed neighbourhood of $x$ is $N[\SG;x]=\{(x_1,x_2), (x_1-1,x_2), (x_1+1,x_2), (x_1,x_2-1), (x_1,x_2+1)\}$ and the corresponding set in $C_n(1,d)$ is  $\{x_1+x_2\cdot d,x_1-1+x_2\cdot d, x_1+1+x_2\cdot d,x_1+(x_2-1)\cdot d, x_1+(x_2+1)\cdot d\}=N[C_n(1,d); x_1+x_2\cdot d]$ (see Figure~\ref{corres}).

We define the following code in the square grid
$$
C_\SG=\{(x_1,x_2)\in  \Z^2 \mid x_1+x_2\cdot d\in C\} \text{.}
$$
In other words, a vertex $(x_1,x_2)$ belongs to $C_\SG$ if and only if the corresponding vertex $x_1+dx_2$ belongs to $C$. In what follows, we show that $C_\SG$ is an identifying code in $\SG$.

Suppose there exist two distinct vertices $x=(x_1,x_2)\in  \Z^2$ and $y=(y_1,y_2)\in  \Z^2$  in the square grid such that $I(\SG;x)=I(\SG;y)$. As $C$ is a dominating set, so is $C_\SG$ and the sets $I(\SG;x)$ and $I(\SG;y)$ are nonempty. Consequently, it suffices to consider the cases where the distance between $x$ and $y$ is at most two in $\SG$. Without loss of generality, we can assume further that the second coordinate of $x$ satisfies $x_2\le y_2$ (if this is not the case, just switch the roles of $x$ and $y$). In other words, $y \in S=\{(x_1,x_2), (x_1+1,x_2), (x_1-1,x_2),(x_1,x_2+1), (x_1+1,x_2+1), (x_1-1,x_2+1), (x_1+2,x_2), (x_1-2,x_2), (x_1,x_2+2)\}$. In the circulant graph $C_n(1,d)$, the property $I(\SG;x)=I(\SG;y)$ implies that $I(C_n(1,d); x_1+x_2\cdot d)= I(C_n(1,d); y_1+y_2\cdot d)$. Because $C$ is identifying, this implies that $x_1+x_2\cdot d\equiv y_1+y_2\cdot d \pmod{n}$. Writing $y_1=x_1+a$ and $y_2=x_2+b$, we obtain $a+b\cdot d\equiv 0 \pmod{n}.$ Notice that the choices for $a$ and $b$ are restricted by $S$. This shows that $I(\SG;x)\neq I(\SG;y)$ in all the other cases (recall that $d\le n/2$, i.e., $n\ge 2d$) except when $y=(x_1,x_2+2)$ and $n=2d.$ Although in this case the sets $I(x_1+x_2\cdot d)$ and $I(y_1+y_2\cdot d)$ are the same in the circulant graph, it is easy to check that the sets $I(\SG;x)$ and $I(\SG;y)$ are not. Indeed, suppose that $y=(x_1,x_2+2)$ and $n=2d$. Notice that $N[\SG;x]\cap N[\SG;y]=\{(x_1,x_2+1)\}.$ If $I(\SG;x)=I(\SG;y)$, the only  codeword in $I(\SG;y)$ can be $(x_1,x_2+1).$ However, in that case there would be also a codeword in $(y_1,y_2+1)$ due to the structure of $C_\SG$ and thus $I(\SG;x)\neq I(\SG;y)$.

For the locating-dominating codes the proof is analogous --- just notice that a non-codeword $x=(x_1,x_2)\in \Z^2$ in $\SG$ corresponds to a non-codeword  $x_1+x_2\cdot d$  in $C_n(1,d)$.

Suppose then that $C$ is self-identifying. We will show that $I(\SG;x)\setminus I(\SG;y)\neq \emptyset$ for all distinct vertices $x=(x_1,x_2)\in \Z^2$ and $y=(y_1,y_2)\in \Z^2$. Since $C_\SG$ is dominating, the claim is clear if $d_\SG(x,y)\ge 3$.

Suppose then that $d_\SG(x,y)=2.$ Denote for any $z=(z_1,z_2)\in \Z^2$ the set $P(z;a,b)=\{z,z+(a,0),z+(0,b)\}$ where  $a,b\in \{-1,1\}$. Let us first observe that $P(z;a,b)$ always contains a codeword of $C_\SG$. This follows since in the circulant graph the set $I(C_n(1,d);z_1+z_2\cdot d)\setminus I(C_n(1,d);z_1-a+(z_2-b)\cdot d)$ contains a codeword of $C$ due to the fact that $C$ is self-identifying. Notice that $z_1+z_2\cdot d$ and $z_1-a+(z_2-b)\cdot d$ are different vertices in $C_n(1,d)$ as $n\ge 2d$. If $y=(x_1-1,x_2-1)$ (resp. $y=(x_1-2,x_2)$), then $N[\SG;x]\setminus N[\SG;y]$ equals $P(x;1,1)$ (resp. contains $P(x;1,1)$). Thus, $I(\SG;x)\setminus I(\SG;y)\neq \emptyset$. Similarly, it is easy to check that for all $x$ and $y$ such that $d_\SG(x,y)=2$, the set $N[\SG;x]\setminus N[\SG;y]$ contains $P(x;a,b)$ for suitable $a,b\in \{-1,1\}$.

Let $d_\SG(x,y)=1$. Similarly as above we can show (looking now at the vertices $x_1+x_2\cdot d$ and $y_1+y_2\cdot d$ in the circulant graph) that the set $N[\SG;x]\setminus N[\SG;y]$ always contains a codeword of $C_\SG$.

%Suppose then that $C$ is self-identifying. We will show that $I(\SG;x)\setminus I(\SG;y)\neq \emptyset$ for all distinct vertices $x=(x_1,x_2)\in \Z^2$ and $y=(y_1,y_2)\in \Z^2$. Since $C_\SG$ is dominating, the claim is clear if $d_\SG(x,y)\ge 3$. Suppose then that $d_\SG(x,y)=2.$ Denote $P(x)=\{x,x+(1,0),x+(0,1)\}.$ If $y=(x_1-1,x_2-1)$ (resp. $y=(x_1-2,x_2)$), then $N[\SG;x]\setminus N[\SG;y]$ equals $P(x)$ (resp. contains $P(x)$). Now $P(x)$ always contains a codeword of $C_\SG$, because in the circulant graph the set $I(C_n(1,d);x_1+x_2\cdot d)\setminus I(C_n(1,d);x_1-1+(x_2-1)\cdot d)$ contains a codeword of $C$ due to the fact that $C$ is self-identifying. Notice that $x_1+x_2\cdot d$ and $x_1-1+(x_2-1)\cdot d$ are different vertices in $C_n(1,d)$ as $n\ge 2d$.
%Similarly, it is easy to check that for all $x$ and $y$ such that $d_\SG(x,y)=2$, the set $N[\SG;x]\setminus N[\SG;y]$ contains $P(x)$ or its rotation around $x$ by an angle of $\pi/2$, $\pi$ or $3\pi/2$ and such a rotation contains a codeword of $C_\SG.$

%For $d_\SG(x,y)=1$, the set $N[\SG;x]\setminus N[\SG;y]$ equals $A(x)=\{x+(1,0),x+(0,1),x+(0,-1)\}$ or its rotation around $x$ by $\pi/2$, $\pi$ or $3\pi/2.$ Similarly as above, we can show that $A(x)$ contains always a codeword of $C_\SG$.
\end{proof}

\begin{figure}[!ht]
%\begin{center}
\include{CnTOGrid}
%\end{center}
\caption{The code $C=\{0,1,2,3,5,7,9,11,13,15\}$ of $C_{17}(1,4)$ on the $2$-dimensional square grid. The crosses mark the codewords and dots the non-codewords.}
\label{corres}
\end{figure}

The previous theorem (together with the results presented in Table~\ref{TableOptimalDensities}) immediately imply the following corollary, which gives lower bounds for the optimal sizes of identifying, locating-dominating and self-identifying codes in the circulant graphs $C_n(1,d)$. Later, in Sections~\ref{SecIDLD} and \ref{SecSID}, we show that the lower bounds can be attained for certain circulant graphs.
\begin{corollary}\label{sqLB}
Let $n$ and $d$ be positive integers such that $ d \geq 2$ and $G = C_n(1,d)$. Then we have
\[
 \LD(G) \geq \left\lceil \frac{3n}{10} \right\rceil \text{, } \ID(G) \geq \left\lceil \frac{7n}{20} \right\rceil  \text{ and } \SID(G) \geq \left\lceil \frac{n}{2} \right\rceil \text{.}
\]
\end{corollary}

In the following theorem, we present the connection between  identifying, locating-dominating and self-identifying codes in the triangular grid and the circulant graphs $C_n(1,d-1,d)$.
\begin{theorem}
Let $n$, $d$ and $k$ be positive integers such that $d\ge 3$. If $C$ is an identifying code in $C_n(1,d-1,d)$ with $k$ codewords, then there exists an identifying code in the infinite triangular grid $\TG$ with density $k/n$. Analogous results also hold for locating-dominating and self-identifying codes.
\end{theorem}
\begin{proof} We take the advantage of the correspondence of a vertex $x=(i,j)$ in the triangular grid $\TG$ and the vertex $i+j\cdot(d-1) \pmod{n}$ in the circulant graph $C_n(1,d-1,d).$ Now the set $N[\TG;x]=\{(i+1,j+1),(i,j+1),(i-1,j),(i,j),(i+1,j),(i,j-1),(i-1,j-1)\}$  corresponds
to the set $N[i+j\cdot(d-1)]$ in the circulant graph. Let $C$ be an identifying  code in
$C_n(1,d-1,d).$ The code  $C_{\TG}=\{(i,j)\in \Z^2\mid i+j\cdot(d-1)\in C\}$ can be shown to be identifying in $\TG$ using similar arguments as in Theorem~\ref{reduction} and the claim follows for identifying codes. %Analogous reasoning gives that if $C$ is locating-dominating (resp. self-identifying), then $C_T$ is also locating-dominating (resp. self-identifying).
Analogous reasoning gives that if $C$ is locating-dominating, then $C_{\TG}$ is also locating-dominating. The case of self-identifying codes is even easier than in the proof Theorem~\ref{reduction}, since it is enough, as discussed in \cite{+koodithl}, to check that there is a codeword of $C_{\TG}$ in the set $N[C_{\TG};x]\setminus N[C_{\TG};y]$ for vertices such that $d(x,y)=1$ (other cases follow from this).
\end{proof}

In the following corollary, we present lower bounds for the circulant graphs $C_n(1,d-1,d)$. In Sections~\ref{SecIDLD} and \ref{SecSID}, we show that the lower bounds can be attained with locating-dominating and self-identifying codes and that there exists an infinite family of identifying codes approaching the lower bound.
\begin{corollary}\label{triLB}
Let $n$ and $d$ be positive integers such that $d\ge 3$ and $G = C_n(1,d-1,d)$. Then we have
\[
\LD(G) \geq \left\lceil \frac{13n}{57} \right\rceil \text{, } \ID(G) \geq \left\lceil \frac{n}{4} \right\rceil \text{ and } \SID(G) \geq \left\lceil \frac{n}{2} \right\rceil \text{.}
\]
\end{corollary}

\medskip

In the following theorem, we present the connection between  identifying, locating-dominating and self-identifying codes in the king grid and the circulant graphs $C_n(1,d-1,d,d+1)$.
\begin{theorem}
Let $n$, $d$ and $k$ be positive integers such that $d\ge 3$. If $C$ is an identifying code in $C_n(1,d-1,d,d+1)$ with $k$ codewords, then there exists an identifying code in the infinite king grid $\KG$ with density $k/n$. Analogous results also hold for locating-dominating and self-identifying codes.
\end{theorem}
\begin{proof} This goes similarly as in  Theorem~\ref{reduction} using the correspondence of a vertex $(i,j)$ in the king grid $\KG$ and the vertex $i+j\cdot d\pmod{n}$ in the circulant graph $C_n(1,d-1,d,d+1)$. The case of self-identifying codes is again easier than in  Theorem~\ref{reduction}, since it suffices, as discussed in \cite{+koodithl}, to check the situation for $d(x,y)=1$ (as other cases follow).
\end{proof}

In the following corollary, we present lower bounds for the circulant graphs $C_n(1,d-1,d,d+1)$. In Sections~\ref{SecIDLD} and \ref{SecSID}, we show that the lower bounds can be attained with locating-dominating and self-identifying codes and that there exists an infinite family of identifying codes approaching the lower bound.
\begin{corollary}\label{kingLB}
Let $n$ and $d$ be positive integers such that $d\ge 3$ and $G = C_n(1,d-1,d,d+1)$. Then we have
\[
\LD(G) \geq \left\lceil \frac{n}{5} \right\rceil \text{, } \ID(G) \geq \left\lceil \frac{2n}{9} \right\rceil \text{ and } \SID(G) \geq \left\lceil \frac{n}{3} \right\rceil \text{.}
\]
\end{corollary}

\section{Identifying and locating-dominating codes in circulant graphs} \label{SecIDLD}

In this section we give optimal constructions for the following types of circulant graphs: $C_n(1,d)$, $C_n(1,d-1,d)$ and $C_n(1,d-1,d,d+1).$

\subsection{On graphs $C_n(1,d)$}

In the next theorem, we will give constructions which attain the bounds in Corollary~\ref{sqLB} for identifying and locating-dominating codes.

\begin{theorem}\label{tightness_1d}
Let $n$ and $d$ be positive integers such that $n \ge  2d$.
\begin{itemize}
\item[(i)]  If $n \equiv 0 \pmod{40}$ and $d \equiv 4 \pmod{40}$, then we have $\ID(C_n(1,d)) = \frac{7n}{20} \text{.}$

\item[(ii)] If $n \equiv 0 \pmod{20}$ and $d \equiv 6 \pmod{20}$, then we have $\ID(C_n(1,d)) = \frac{7n}{20} \text{.}$

%    \]
\item[(iii)] If $n \equiv 0 \pmod{20}$ and $d \equiv 5 \pmod{20}$, then we have $ \LD(C_n(1,d)) = \frac{3n}{10} \text{.}$
%    \]
\end{itemize}
\end{theorem}
\begin{proof}
(i) Let $n \equiv 0 \pmod{40}$ and $d \equiv 4 \pmod{40}$. Define $$B_1 = \{0, 1, 2, 8,  10, 12, 16, 18, 22, 24, 26, 32, 33, 34\}$$ and
\[
D_1 = \{ u \in \Z_n \ | \ u \equiv b \pmod{40} \text{ for some } b \in B_1 \} \text{.}
\]
The codes $B_1$ in $C_{40}(1,4)$ and $D_1$ in $C_{n}(1,d)$, where $n=80$ and $d=44$ are illustrated in Figure~\ref{idTightness}.

\begin{figure}%[!ht]
\begin{tikzpicture}[scale=0.18,rotate=-90,>=triangle 45]
\foreach \x in {3,4,5,6,7,9,11,13,14,15,17,19,20,21,23,25,27,28,29,30,31,35,36,37,38,39,43,44,45,46,47,49,51,53,54,55,57,59,60,61,63,65,67,68,69,70,71,75,76,77,78,79}
\fill[shift={(0,\x)},color=black] circle (5pt);
\foreach \x in
{0,5,10,15,20,25,30,35,40,45,50,55,60,65,70,75,79}
\draw[shift={(-0.1,\x)},color=black] node[below] {\scriptsize $\x$};
\draw[shift={(2,1)},color=black] node[below] {\scriptsize $C_{80}(1,44)$};
\foreach \x in 
{23,24,25,26,27,29,31,33,34,35,37,39,40,41,43,45,47,48,49,50,51,55,56,57,58,59}
\fill[shift={(-4,\x)},color=black] circle (5pt);
\foreach \x in 
{0,5,10,15,20,25,30,35,39}
\draw[shift={(-3.9,20+\x)},color=black] node[above] {\scriptsize $\x$};
\draw[shift={(-6,21)},color=black] node [above] {\scriptsize $C_{40}(1,4)$};
\foreach \x in 
{0,1,2,8,10,12,16,18,22,24,26,32,33,34,40,41,42,48,50,52,56,58,62,64,66,72,73,74}
\draw [shift={(0,\x)},color=black] (0,0)-- ++(-7.0pt,-7.0pt) -- ++(14pt,14.0pt) ++(-14.0pt,0) -- ++(14.0pt,-14.0pt);
\foreach \x in 
{20+0,20+1,20+2,20+8,20+10,20+12,20+16,20+18,20+22,20+24,20+26,20+32,20+33,20+34}
\draw [shift={(-4,\x)},color=black] (0,0)-- ++(-7.0pt,-7.0pt) -- ++(14pt,14.0pt) ++(-14.0pt,0) -- ++(14.0pt,-14.0pt);
\draw (-5.5,19)--(-2.5,19)--(-2.5,60)--(-5.5,60)--(-5.5,19);
\draw (-1.5,-1)--(1.5,-1)--(1.5,80)--(-1.5,80)--(-1.5,-1);
\draw [->,color=black] (-4,20) -- (-2,10);
\draw [color=black] (0,0) -- (-2,10);
\draw [->,color=black] (-4,20) -- (-2,30);
\draw [color=black] (0,40) -- (-2,30);
\draw [->,color=black] (-4,59) -- (-2,49);
\draw [color=black] (0,39) -- (-2,49);
\draw [->,color=black] (-4,59) -- (-2,69);
\draw [color=black] (0,79) -- (-2,69);
\end{tikzpicture}
\caption{Optimal identifying codes for $C_{40}(1,4)$ and $C_{80}(1,44)$. The crosses denote the codewords and dots are non-codewords.}
  \label{idTightness}
\end{figure}
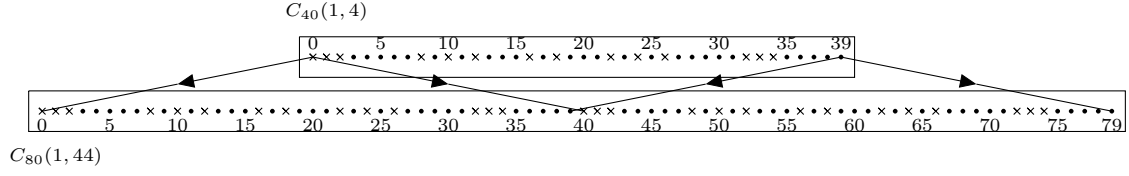

It is straightforward to verify that $B_1$ is an identifying code in $C_{40}(1,4)$. In what follows, we prove that $D_1$ is an identifying code in $C_{n}(1,d)$ by showing that all the identifying sets $I(C_{n}(1,d),D_1; x)$ are nonempty and unique. Observe first that by the construction of $D_1$ we obtain for all $x \in \Z_n$ that
\[
I(C_{n}(1,d),D_1;x) \equiv I(C_{40}(1,4),B_1;x') \pmod{40} \text{,}
\]
where $x'$ is an integer such that $x \equiv x' \pmod{40}$ and $0 \leq x' \leq 39$. Therefore, the identifying sets $I(C_{n}(1,d),D_1;x)$ are nonempty for all $x \in \Z_n$. Let $x$ and $y$ be distinct vertices of $\Z_n$. Assume first that $x \not\equiv y \pmod{40}$. Let then $x'$ and $y'$ be integers such that $x \equiv x' \pmod{40}$, $y \equiv y' \pmod{40}$, $0 \leq x' \leq 39$ and $0 \leq y' \leq 39$. %Notice first that by the construction of $D_1$ we have
%\[
%I(C_{n}(1,d),D_1;x) \equiv I(C_{40}(1,4);B_1,x') \pmod{40}
%\]
%and
%\[
%I(C_{n}(1,d),D_1;y) \equiv I(C_{40}(1,4);B_1,y') \pmod{40} \text{.}
%\]
Therefore, by the previous observation, if $I(C_{n}(1,d),D_1;x) = I(C_{n}(1,d),D_1;y)$, then $I(C_{40}(1,4);B_1,x') = I(C_{40}(1,4);B_1,y')$ and we have a contradiction as $B_1$ is an identifying code in $C_{40}(1,4)$. Hence, we may assume that $x \equiv y \pmod{40}$. Let us then show that $N[C_{n}(1,d);x] \cap N[C_{n}(1,d);y] = \emptyset$. Suppose to the contrary that there exist $x, y \in \Z_n$ such that $x+j = y+j'$ for some $j, j' \in \{-d,-1,0,1,d\}$. Since $x \equiv y \pmod{40}$, we obtain that $j \equiv j' \pmod{40}$. This further implies that $j = j'$ and $x = y$ (a contradiction). %Then it is straightforward to verify that $N[C_{n}(1,d);x] \cap N[C_{n}(1,d);y] = \emptyset$.
Therefore, as each vertex of $\Z_n$ is covered by a codeword of $D_1$, we have $I(C_{n}(1,d),D_1;x) \neq I(C_{n}(1,d),D_1;y)$. Thus, $D_1$ is an identifying code in $C_{n}(1,d)$.

(ii) Let $n \equiv 0 \pmod{20}$ and $d \equiv 6 \pmod{20}$. Define $B_2 = \{0, 2, 8, 9, 11, 12, 18\}$ and
\[
D_2 = \{ u \in \Z_n \ | \ u \equiv b \pmod{20} \text{ for some } b \in B_2 \} \text{.}
\]
It is straightforward to verify that $B_2$ is an identifying code in $C_{20}(1,6)$. Then, using similar arguments as in the case~(i), we can prove that $D_2$ is an identifying code in $C_n(1,d)$.

(iii) Let $n \equiv 0 \pmod{20}$ and $d \equiv 5 \pmod{20}$. Define $B_3 = \{0, 4, 7, 11, 14, 17\}$ and
\[
D_3 = \{ u \in \Z_n \ | \ u \equiv b \pmod{20} \text{ for some } b \in B_3 \} \text{.}
\]
It is straightforward to verify that $B_3$ is a locating-dominating code in $C_{20}(1,5)$. Then, using similar arguments as in the case~(i) (although now $x$ and $y$ are assumed to be non-codewords), we can prove that $D_3$ is a locating-dominating code in $C_n(1,d)$.
\end{proof}

\subsection{On graphs $C_n(1,d-1,d)$}

The next theorem gives optimal constructions on locating-dominating codes in $C_n(1,d-1,d)$.  In addition, we provide an infinite sequence of identifying codes approaching the lower bound  for identifying codes in Corollary~\ref{triLB}. Moreover, it will be shown  in Corollary~\ref{CorollaryUnreachable} (see also Theorem~\ref{nonreachableThm}) that  we cannot attain the lower bound by any identifying code.

\begin{theorem}\label{Triangular_tightness}
\begin{itemize}
\item[(i)]
 For all the parameters $n$ and $d$ such that $d\equiv 8 \pmod{57}$, $d\ge 8 $, $n\ge 2d$ and $n\equiv 0 \pmod{57}$, we have $\LD(C_n(1,d-1,d)) = \frac{13n}{57} \text{.}$

\item[(ii)] We have a sequence of identifying
codes $(C_k)_{k=1}^{\infty}$ in the circulant graphs $C_n(1,d-1,d)$
 with
$$\lim_{k\rightarrow \infty} \frac{|C_k|}{n}=1/4. $$
\end{itemize}
\end{theorem}
\begin{proof}
(i) Let $d\equiv 8 \pmod{57}$, $d\ge 8 $, $n\ge 2d$ and $n\equiv 0 \pmod{57}$. We denote
$$
B=\{0,2,4,6,15,18,27,29,31,33,43,45,47\}.
$$
Let further
$$
C=\{v\in \Z_n \mid v\equiv b \pmod{57} \textrm{ for some } b\in B\}.
$$
It is straightforward to check that $B$ is a locating-dominating code in $C_{57}(1,d-1,d)$ for $d=8.$ Next we will show that $C$ is locating-dominating in $C_n(1,d-1,d)$. Let us first show that $I(x)=I(y)$ for $x\not\equiv y\pmod{57}$ and $x,y\notin C$. Denote $x'=x\pmod{57}$ and $y'= y\pmod{57}$ where $0\le x'\le 56$ and $0\le y'\le 56$. If $I(x)=I(y)$, then it follows that the codewords in $I(x)$ and in $I(y)$ would be equal modulo 57. However, that is not possible, since $I(B; x')\neq I(B;y')$ for distinct $x',y'\notin B.$ Therefore, it suffices to consider $I(x)=I(y)$ for $x\equiv y\pmod{57}$, $x\neq y$ and $x,y\notin C$.
Let $j\in\{-d,d+1,-1,0,1,d-1,d\}$ and  $x+j\in I(x)$. Consequently, $x+j=y+j'$ for some $j'\in\{-d,d+1,-1,0,1,d-1,d\}.$ Since $x\equiv y\pmod{57}$, we get $j=j'$ giving $x=y.$
 Hence $C$ is locating-dominating and it attains the lower bound in Corollary~\ref{triLB}.

\medskip

(ii) Let $d\ge 6$ be even and $n=6d$. Denote  $S=\{j\mid 0\le j \le d, j\equiv 0\pmod{2}\}.$ We define
$$
C_d=\{v\in \Z_n \mid v\equiv b \pmod{2d} \textrm{ for some } b\in S\}.
$$
The code $C_d$ has  cardinality  $3(d/2+1)$. Thus
$\lim_{d\rightarrow \infty}|C_d|/n=1/4.$

We will show that $C_d$ is identifying in $C_n(1,d-1,d)$.
If $x\equiv s \pmod{2d}$ with $d\le s\le 2d-1$ and $x$ is odd, then $\{x-d+1,x+d-1\}\subseteq I(x)$. Since $N[x-d+1]\cap N[x+d-1]=\{x\}$, it follows that $I(x)\neq I(y)$ for any $y\neq x $.  If $x\equiv s \pmod{2d}$ where $x$ is even and  $d\le s\le 2d-1$ or $s=0$, then   $\{x-d,x+d\}\subseteq I(x)$. Since $N[x-d]\cap N[x+d]=\{x\}$, the $I(x)$ is distinguished from other $I(y)$'s. Suppose then that $x\equiv s \pmod{2d}$ with $1\le s\le d-1$ and $x$ is odd. Now $\{x-1,x+1\}\subseteq I(x)$ and again $I(x)$ is unique among $I$-sets. If $x\equiv s \pmod{2d}$ with $1\le s\le d-1$ and $x$ is even, then $I(x)=\{x\}$. It follows that $C_d$ is identifying.
\end{proof}

\subsection{On graphs $C_n(1,d-1,d,d+1)$}
In the following theorem, we give optimal  locating-dominating  codes in the circulant graph $C_n(1,d-1,d,d+1)$. Furthermore,  we  give  an infinite sequence of identifying codes approaching the lower bound in Corollary~\ref{kingLB}.

\begin{theorem} \label{KingThm}
\begin{itemize}

\item[(i)] For $d\equiv 8 \pmod{10}$, %or $d\equiv 2 \pmod{10}$,
 $d\ge 8$, $n\ge 4d+6$ and $n\equiv 0
\pmod{10}$, we have $ \LD(C_n(1,d-1,d,d+1)) = \frac{n}{5} \text{.}$

\item[(ii)] There is a  sequence of
identifying codes $(C_k)_{k=1}^{\infty}$ in the circulant graphs $C_n(1,d-1,d,d+1)$
 with
$$\lim_{k\rightarrow \infty} \frac{|C_k|}{n}=2/9. $$
\end{itemize}
\end{theorem}

\begin{proof}
(i) Let $d\equiv 8\pmod{10}$, $n\ge 4d+6$ and $n\equiv 0 \pmod{10}.$ Next we will verify that the code
$$C'=\{v\in  \Z_n \mid v\equiv 0,4 \pmod{10}\}$$
is locating-dominating in $C_n(1,d-1,d,d+1)$. Notice that the size of $C'$ attains the lower bound in Corollary~\ref{kingLB}.
Since $d\equiv 8 \pmod{10}$, then we get the following $I$-sets depending on the value of non-codewords $x$ modulo 10
$$\begin{array}{c|c|c}
  x \pmod{10} & I(x) & I(x)\pmod{10} \\ \hline
  1 & \{x-1,x-d+1,x+d+1\} & 0,4,0\\
  2 & \{x-d,x+d\} & 4,0\\
  3 & \{x+1,x-d-1,x+d-1\} & 4,4,0 \\
  5 & \{x-1,x+d+1\} & 4,4\\
  6 & \{x+d\} & 4\\
  7 & \{x-d+1,x+d-1\} & 0,4\\
  8 & \{x-d\} & 0\\
  9 & \{x+1,x-d-1\} & 0,0.
\end{array}
$$
Let $x\neq y$.
Clearly,  $I(x)\neq I(y)$ for those $x$ and $y$ which have different sizes of the $I$-sets. Let us first consider the cases where the size of the $I$-sets equal one. If $x\equiv 6 \pmod{10}$ and $y\equiv 8 \pmod{10}$, then (see the table above) $c\in I(x)$ has $c\equiv 4 \pmod{10}$ and $c'\in I(y)$ has $c'\equiv 0\pmod{10}.$ Therefore, $I(x)\neq I(y)$. Obviously, the sets $I(x)\neq I(y)$ if $x\equiv y\equiv 6 \pmod{10}$ or if $x\equiv y\equiv 8 \pmod{10}$. Consider then the case of $I$-sets of size three. Let first $x\equiv 1 \pmod{10}$ and $y\equiv 3 \pmod{10}$. Now the set $I(x)$ has exactly one codeword $c$ such that $c\equiv 4\pmod{10}$ and the set $I(y)$ has exactly two such codewords. Therefore, $I(x)\neq I(y)$. Consider then
the case  $x\equiv y\equiv 1 \pmod{10}$. Now the only codeword which is 4 modulo 10 is $x-d+1$ in $I(x)$ and $y-d+1$ in $I(y).$ Consequently, if $I(x)=I(y)$, then $x-d+1\equiv y-d+1\pmod{n}$ giving $x=y$ (in $\Z_n$). The case if $x\equiv y\equiv 3 \pmod{10}$ goes similarly.
Consider then the $I$-sets of size two. We start with the situation $I(x)=I(y)$ where $x \not\equiv y \pmod{10}.$ If $x\equiv 5\pmod{10}$ (resp. $x\equiv 9 \pmod{10})$, then in $I(x)$ both of the codewords are equal to 4 (resp. 0) modulo 10. If $x\equiv 2 \pmod{10}$ or $x\equiv 7 \pmod{10}$, then the $I(x)$ has exactly one codeword 0 modulo 10 and one 4 modulo 10. Therefore, it suffices to consider the case
 $x\equiv 2 \pmod{10}$ or $y\equiv 7 \pmod{10}$. Now $I(x)=\{x-d,x+d\}$ and $I(y)=\{y-d+1, y+d-1\}.$ Due to the residue classes modulo 10, we must have $x-d\equiv y+d-1\pmod{n}$ and $x+d\equiv y-d+1\pmod{n}$. This implies that $2x\equiv 2y \pmod{n}$. If $n$ is odd, we immediately have $x=y$ (in $\Z_n$). If $n$ is even, we still have $x=y$ due to the fact that  $n\ge 4d+6.$

  The cases $x \equiv y\equiv 2 \pmod{10}$ and $x \equiv y\equiv 7 \pmod{10}$ go as above based on the residue classes modulo 10 of the codewords in $I(x)$ and $I(y)$. In the cases $x \equiv y\equiv 5,9 \pmod{10}$  we use the fact that $n\ge 4d+6.$ In summary $I(x)\neq I(y)$ for $x\neq y$ we we obtain the assertion.

\medskip

(ii) The proof is somewhat technical and postponed to the Appendix.
\end{proof}

\section{Self-identifying codes in circulant graphs} \label{SecSID}

In the next theorem, we will show that the bounds on self-identifying codes in  Corollaries \ref{sqLB}, \ref{triLB} and \ref{kingLB} can be reached.
%\begin{theorem}
%\begin{itemize}
%\item[(i)] Let  $d\ge 4$, $d$ be even, $n\ge 4d+1$ and $n\equiv 0 \pmod{2}$. For all of these %parameters $ \SID(C_n(1,d)) = \frac{n}{2} \text{.}$
%\item[(ii)] Let $d\ge 4$.
%For the parameters   $d\ge 4$, $n\ge 4d+6$ and $n\equiv 0 \pmod{2}$ we have $\SID(C_n(1,d-1,d)) = %\frac{n}{2} \text{.}$
%\item[(iii)] Let $d\ge 4$.
%If $d\equiv 1 \pmod{3}$,
%$d\ge 4$, $n\ge 3d+5$ and $n\equiv 0 \pmod{3}$, then $ \SID(C_n(1,d-1,d,d+1)) = \frac{n}{3} %\text{.}$
%\end{itemize}
%\end{theorem}

\begin{theorem} \label{ThmSIDConstructions}
Let $d$ be an integer such that $d \geq 4$.
  \begin{itemize}
\item[(i)] If $d$ is even, $n\ge 4d+1$ and $n\equiv 0 \pmod{2}$, then we have $\SID(C_n(1,d)) = \frac{n}{2}$.
\item[(ii)] If $n\ge 4d+1$ and $n\equiv 0 \pmod{2}$, then we have $\SID(C_n(1,d-1,d)) = \frac{n}{2}$.
\item[(iii)] If $d\equiv 1 \pmod{3}$, $n\ge 4d+5$ and $n\equiv 0 \pmod{3}$, then $\SID(C_n(1,d-1,d,d+1)) = \frac{n}{3}$.
\end{itemize}
\end{theorem}
\begin{proof} (i) We show that the code
$$
C=\{v\in \Z_n \mid v\equiv 0\pmod{2}\}
$$
is self-identifying in the circulant graph $C_n(1,d)$. If $x\equiv 0\pmod{2}$, then $I(x)=\{x-d,x,x+d\}$ and otherwise $I(x)=\{x-1,x+1\}.$ Since $n\ge 4d+1$, we get that $N[x-d]\cap N[x-d]=\{x\}$ and $N[x-1]\cap N[x+1]=\{x\}.$ Consequently, the condition for self-identification, namely, $\cap_{c\in I(x)}N[c]=\{x\}$, is satisfied. As $\frac{n}{2}$ is the lower bound, we showed that $\SID(C_n(1,d))=\frac{n}{2}$.
%\begin{figure}[!h]
 % \caption{Optimal SID-code for $C_{34}(1,8)$}
  %\label{tightnessn2}
  %\include{figures/tightnessn2}
  %\end{figure}

\medskip

(ii) Let $d\ge 4$, $n\ge 4d+1$ and $n$ be even. The code
$$C=\{v\in  \Z_n\mid v\equiv 0\pmod{2}\}$$
is self-identifying in $C_n(1,d-1,d)$ as will be seen next.
If $d$ is even (resp. odd) and $x\equiv 0\pmod{2}$, then $\{x-d,x+d\}\subseteq I(x)$ (resp. $\{x-d+1,x+d-1\}\subseteq I(x)$). Hence in both cases
$\cap_{c\in I(X)}N[c]=\{x\}.$ If $d$ is even (resp. odd) and $x\equiv 1 \pmod{2}$, then $\{x-d+1,x+d-1\}\subseteq I(x)$ (resp. $\{x-d,x+d\}\subseteq I(x)$). Consequently, again $\cap_{c\in I(x)}N[c]=\{x\}.$ Therefore, $C$ is self-identifying. As $\frac{n}{2}$ is the lower bound, we showed that $\SID(C_n(1,d-1,d))=\frac{n}{2}$.

\medskip

(iii) Let
$$C=\{v\in  \Z_n \mid v\equiv 0 \pmod{3}\}.$$ We verify next that $C$ is
self-identifying in $C_n(1,d-1,d,d+1)$. If $x\equiv 0 \pmod{3}$, we have
$I(x)=\{x,x-d+1,x+d-1\}$ since $d\equiv 1 \pmod{3}$. If $x\equiv 1 \pmod{3}$ (resp. $x\equiv 2\pmod{3})$, then
$I(x)=\{x-1,x-d,x+d+1\}$ (resp. $I(x)=\{x+1,x-d-1,x+d\})$.
Now in each case, the intersection $\cap_{c\in I(x)}N[x]=\{x\}$ due to the fact that $n\ge 4d+5.$ Hence $C$ is self-identifying.
\end{proof}

In what follows, we give the optimal cardinalities of self-identifying codes in $C_n(1,3)$ and $C_n(1,4)$ (for $n$ odd). In these cases, the optimal cardinalities do \emph{not} attain the $n/2$ lower bound of Corollary~\ref{sqLB}, and for this purpose, we introduce new methods for increasing the lower bounds. In the following proposition, we present some results which are useful in the upcoming proofs.
\begin{proposition}\label{SymmetricSIDnonC}
Let $n$ and $d_1< d_2$ be integers such that $4d_2-1<n$.
If $K$ is a self-identifying code in $C_n(d_1,d_2)$, then the following statements hold:
\begin{itemize}
 \item[(i)] For all $x\in K$, we have $|I(x)|> 2$.
  \item[(ii)] For all $x\notin K$, there exists $1\leq i\leq 2$ such that  $\{x-d_i,x+d_i\}\subseteq I(x)$. % and $|I(x)|>1$.
  \item[(iii)] If $d_1=1$, $d_2=3$ and $|I(x)|=2$, then we have  $I(x)=\{x-3,x+3\}$ for all $x\notin K$.
\end{itemize}
\end{proposition}
\begin{proof} Let $x$ be a vertex in the code. Assume it has only two vertices: itself and $y$. Then $I(y)$ contains the same two vertices. Hence, $I(x)$ contains at least three vertices.

Let $x$ then be a non-codeword. Assume that $I(x)$ does not contain the claimed subset. Then, without loss of generality, we can assume that either $I(x)=\{x-d_1,x+d_2\}$ or $I(x)=\{x+d_1,x+d_2\}.$  Suppose first that $I(x)=\{x-d_1,x+d_2\}$. For $y=x-d_1+d_2$ we have $I(y)=\{y-d_2,y-d_1,y,y+d_1,y+d_2\}\cap K$. Now $y-d_2=x-d_1$ and $y+d_1=x+d_2$ are both in $I(y)$ giving $I(x)\subseteq I(y)$. Thus, $K$ is not self-identifying. Assume then that $I(x)=\{x+d_1,x+d_2\}$. Now if $y=x+d_1+d_2$, then as above $I(x)\subseteq I(y)$ and we are done.
%If  $x-d_1=x+d_1$ then to be well identified $I(x)$ contains $\{x-d_2,x+d_2\}$ and vice versa. Otherwise $I(x-d_1)$ contains $I(x)$.
For $d_1=1$ and $d_2=3$ we cannot have $I(x)=\{x-1,x+1\}$ since  $N[x-1]\cap N[x+1]=\{x,x-2,x+2\}.$
\end{proof}

In the following theorem, we present the sizes of optimal self-identifying codes in $C_n(1,3)$ for all integers $n>11$. In particular, we show that any self-identifying code in $C_n(1,3)$ has at least $\lceil 4n/7 \rceil$ codewords and that there exists a construction attaining this lower bound when $n \equiv 0, 1, 2, 4, \text{or } 6 \pmod{7}$. In the cases $n \equiv 3 \pmod{7}$ and $n \equiv 5 \pmod{7}$, we increase the lower bound by one using a novel technique and present constructions meeting this improved lower bound.
\begin{theorem} \label{Optim13} %EUROCOMB
The optimal cardinalities of self-identifying codes in $C_n(1,3)$ for $n>11$ are as follows:
$$\SID(C_n(1,3))=\left\{ \begin{array}{ll}
                             4k & \textrm{if } n=7k \\
                             4k+1 & \textrm{if }  n=7k+1 \\
                             4k+2 & \textrm{if }  n=7k+2 \\
                             4k+3 & \textrm{if } n\in \{7k+3,7k+4\} \\
                             4k+4 & \textrm{if } n\in \{7k+5,7k+6\}
                           \end{array}
\right. .$$
 % $$
%  \begin{array}{llll}
%  n=7k&  \SID(C_n(1,3)) & = & 4k \\
%  n=7k+1&  \SID(C_n(1,3)) & = & 4k+1\\
%  n=7k+2&  \SID(C_n(1,3)) & = & 4k+2\\
%  n\in \{7k+3,7k+4\}&  \SID(C_n(1,3)) & = & 4k+3\\
%  n\in \{7k+5,7k+6\} & \SID(C_n(1,3)) & = & 4k+4.\\
%
%    \end{array}
%  $$
\end{theorem}
\begin{proof}
Let $n$ be an integer such that $n>11$. Observe first that we have the following characterization for self-identifying codes in $C_n(1,3)$:
\begin{itemize}
\item A code $K$ in $C_n(1,3)$ is self-identifying if and only if $|I(K;c)| \geq 3$ for all $c \in K$ and $\{u-3, u+3\} \subseteq I(K;u)$ for all $u \in \Z_n \setminus K$.
\end{itemize}
Indeed, if $K$ is a self-identifying code in $C_n(1,3)$, then the given conditions are met by the previous proposition. On the other hand, if $K$ satisfies the conditions, then it is straightforward to verify that $K$ is a self-identifying code by the characterization~\eqref{CharacterizationSID}.

Let $K$ be a self-identifying code in $C_n(1,3)$. In what follows, we study more closely what happens if there exists consecutive non-codewords in $K$:
\begin{itemize}
\item If there are four or more non-consecutive non-codewords, then the first one, say $u$, contradicts with the previous characterization as $u+3$ does not belong to $K$.
\item If there are exactly three consecutive non-codewords, say $\{0,1,2\}$ (and thus $n-1$ and $3$ are in the code), then $\{n-4,n-3,n-2,4,5,6\}$ are all codewords (by the characterization). Let $P3$ be the pattern with $3$ consecutive non-codewords followed by four consecutive codewords (see Figure~\ref{patternsC13}).
\item If there are exactly two consecutive non-codewords, say $\{0,1\}$, then $\{n-3,n-2,n-1,2,3,4\}$ are in the code. Let $P2$ be the pattern with two consecutive non-codewords followed by three consecutive codewords as in Figure~\ref{patternsC13}.
\item Suppose then that there is only one consecutive codeword, say non-codeword 0 (and $n-1$ and $1$ are in the code). If $2\in K$, then we get the pattern $P1a$ with one non-codeword followed by two codewords. On the other hand, if $2\notin K$, then we obtain (by the characterization) the pattern  $P1b$ with five consecutive vertices with only the first and the third one being non-codewords.
\end{itemize}
Notice that the smallest density among the patterns is the one with three consecutive non-codewords followed by four codewords, i.e., the density of the codewords in the patter is $4/7$.
%Let $n$ be an integer $n>11$, and $K$ a self-identifying code in $C_n(1,3)$.  We consider neighbourhoods of consecutive non-codewords using the properties of Proposition~\ref{SymmetricSIDnonC}.

%If there are four consecutive non-codewords, the first and last ones are not self-identified according to (iii) in Proposition~\ref{SymmetricSIDnonC}. If there are three consecutive non-codewords, say $\{0,1,2\}$ (and thus $n-1$ and $3$ are in the code), then $\{n-4,n-3,n-2,4,5,6\}$ are all codewords (by (i) and (iii) of Proposition \ref{SymmetricSIDnonC}). Let $P3$ be the pattern with $3$ consecutive non-codewords followed by four consecutive codewords (see Figure~\ref{patternsC13}). If there are two consecutive non-codewords, say $\{0,1\}$, then $\{n-3,n-2,n-1,2,3,4\}$ are in the code. Let $P2$ be the pattern with two consecutive non-codewords followed by three consecutive codewords as in Figure~\ref{patternsC13}. Assume then that there is only one non-codeword 0 (and $n-1$ and $1$ are in the code). If $2\in K$, then we get the pattern $P1a$ with one non-codeword followed by two codewords and, if $2\notin K$, then we obtain (using Proposition~\ref{SymmetricSIDnonC}) the pattern  $P1b$ with five consecutive vertices with only the first and the third non-codewords. The smallest density among the patterns is the one with three consecutive non-codewords (i.e. $4/7$).

Due to the obtained patterns, we may conclude that there exists in the graph two consecutive codewords followed by a non-codeword. Without loss of generality, it can be assumed that $n-2, n-1 \in K$ and $0 \notin K$. %Assume the vertices $n-2$ and $n-1$ are both in the code and $0$ is not in the code.
%We can assume this without loss of generality since there must be at least one non-codeword in optimal self-identifying code, every pattern has at least two consecutive codewords and we can shift the names on the vertices.
Furthermore, there exists a vertex $x_1$ such that the set $s_1=\{0,1,\dots,x_1\}$ is one of the patterns $P3$, $P2$, $P1a$ or $P1b$. Hence $x_1-1$ and $x_1$ are codewords and we can do the same thing with the next non-codeword vertex $x_2$ (notice that $x_2$ may be different from $x_1+1$). Let $x_3$ be such that $s_2=\{x_2,x_2+1,\dots,x_3\}$ is one of the patterns. We can go on to the right and define all the sets $s_1,\dots,s_r$ that correspond to the patterns. Note that the vertices that are not in these sets are all codewords. This partition the graph in patterns with maybe some codewords separating them. Notice also that the last pattern $s_r$ do not intersect the first one $s_1$.  For each of these sets $s_i$ let $d_i$ be its density and $n_i$ the number of vertices. The density of $K$ can then be estimated
$$  d\geq \frac1n (\sum_{1\leq i \leq r} d_in_i+n-\sum_{1\leq i\leq r}n_i)\geq \frac1n (\sum_{1\leq i \leq r} \frac47 n_i+n-\sum_{1\leq i\leq r}n_i)=\frac47$$
This implies that the self-identifying code $K$ has at least $\lceil 4n/7 \rceil$ codewords. The proof now divides into the following cases depending on the remainder of $n$ when divided by $7$:
\begin{figure}%[!ht]
  \begin{center}
  \begin{tikzpicture}[scale=0.4]
\clip(-2.7,-0.9) rectangle (12.5,6.3);
	%P3
  \foreach \x in {0,2,4}
  \fill [shift={(\x,6)}] circle (5pt);
  \foreach \x in {6,8,10,12}
  \draw [shift={(\x,6)},color=black] (0,0)-- ++(-7.0pt,-7.0pt) -- ++(14pt,14.0pt) ++(-14.0pt,0) -- ++(14.0pt,-14.0pt);
  \foreach \x in {0,1,2,3,4,5,6}
  \draw [shift={(2*\x,6)}] node [below] {$\x$};
  \draw (-2,6) node {$P3$};
  
	%P2
  \foreach \x in {0,2}
  \fill [shift={(\x,4)}] circle (5pt);
  \foreach \x in {4,6,8}
  \draw [shift={(\x,4)},color=black] (0,0)-- ++(-7.0pt,-7.0pt) -- ++(14pt,14.0pt) ++(-14.0pt,0) -- ++(14.0pt,-14.0pt);
  \foreach \x in {0,1,2,3,4}
  \draw [shift={(2*\x,4)}] node [below] {$\x$};
  \draw (-2,4) node {$P2$};

	%P1a
  \foreach \x in {0}
  \fill [shift={(\x,2)}] circle (5pt);
  \foreach \x in {2,4}
  \draw [shift={(\x,2)},color=black] (0,0)-- ++(-7.0pt,-7.0pt) -- ++(14pt,14.0pt) ++(-14.0pt,0) -- ++(14.0pt,-14.0pt);
  \foreach \x in {0,1,2}
  \draw [shift={(2*\x,2)}] node [below] {$\x$};
  \draw (-2,2) node {$P1a$};

	%P1b
  \foreach \x in {0,4}
  \fill [shift={(\x,0)}] circle (5pt);
  \foreach \x in {2,6,8}
  \draw [shift={(\x,0)},color=black] (0,0)-- ++(-7.0pt,-7.0pt) -- ++(14pt,14.0pt) ++(-14.0pt,0) -- ++(14.0pt,-14.0pt);
  \foreach \x in {0,1,2,3,4}
  \draw [shift={(2*\x,0)}] node [below] {$\x$};
  \draw (-2,0) node {$P1b$};

 \end{tikzpicture}	
  \end{center}
    \caption{The patterns for $C_n(1,3)$. The crosses denote the codewords.}\label{patternsC13}
  \end{figure}
\begin{itemize}
\item If $n=7k$, then the code has at least $\lceil\frac47n\rceil$ codewords, that is, $4k$. The code $K_1=\{i+7j\mid 0\leq i \leq3 , 0\leq j \leq k-1\}$ is self-identifying (see the case $C_{14}(1,3)$ in Figure~\ref{constructionsC13}). Indeed, for every vertex $v\notin K$, we have $\{v-3,v+3\}\subseteq I(v)$. Furthermore, for every vertex $v\in K$, we have $|I(K_1;v)| \geq 3$. Thus, according to the characterization, the code $K_1$ is self-identifying in $C_{n}(1,3)$. %either the vertices $v-1,v$  are in $I(v)$ or the vertices $v,v+1$ are; moreover $I(v)$ contains also one more vertex. These three codewords intersect uniquely in $v$.
\item If $n=7k+1$, then the code has at least $\lceil\frac47n\rceil$ codewords, that is, $4k+1$. By the same argument as for the case $n=7k$, the code $K_2=\{i+7j\mid 0\leq i \leq3 , 0\leq j \leq k-1\}\cup \{7k\}$ can be shown to be self-identifying (see the case $C_{15}(1,3)$ in Figure~\ref{constructionsC13}).
\item If $n=7k+2$, then the code has at least $\lceil\frac47n\rceil=4k+2$ codewords. By the same argument as for the case $n=7k$ the code  $K_3=\{i+7j\mid 0\leq i \leq3 , 0\leq j \leq k-1\}\cup \{7k,7k+1\}$ works (see the case $C_{16}(1,3)$ in  Figure~\ref{constructionsC13}).
\item If $n=7k+4$ (notice that the more difficult case of $n=7k+3$ will be dealt later), then the code has at least $4k+3$ codewords, the code $K_5=\{i+7j\mid 0\leq i \leq3 , 0\leq j \leq k-1\}\cup\{7k-1,7k,7k+1\}$ works (see the case $C_{18}(1,3)$ in Figure~\ref{constructionsC13}). Indeed, as above, it is straightforward to verify that $\{v-3,v+3\} \subseteq I(K_5;v)$ for all $v \notin K$ and $|I(K_5;c)| \geq 3$ for all $c \in K_5$. Thus, $K_5$ is self-identifying by the characterization. %for every vertex $v\notin K$, the vertices $v-3,v+3$ are in $I(v)$ and we are done. For every vertex $v\in K$, the set $I(v)$ has the vertices $v\pm 1,v$ and another one which makes them well self-identified (see the case $C_{18}(1,3)$ in Figure~\ref{constructionsC13}).
\item If $n=7k+6$ (notice that the case $n=7k+5$ is postponed), then the code has at least $4k+4$ codewords. As above, we can show that the code $K_7=\{i+7j\mid 0\leq i \leq3 , 0\leq j \leq k\}$ is self-identifying in $C_{n}(1,3)$ (see the case $C_{13}(1,3)$ in  Figure~\ref{constructionsC13}).
\item Suppose $n=7k+3$. We will first show that now a self-identifying code has at least $4k+3$ codewords. Every self-identifying code on $C_{7k+3}(1,3)$ needs at least $\lceil\frac47n\rceil = 4k+2$ codewords. Assume that there is a self-identifying code $K$ on $C_n(1,3)$ with $4k+2$ codewords. Recall that the density of codewords in the patterns is at least $3/5$ unless the pattern is $P3$. If there are at most $k-2$ patterns of $P3$, then $|K|\ge \frac47 (7(k-2))+\frac35 (n-7(k-2))=4k+\frac{11}{5}>4k+2$. Consequently, there must be either $k$ or $k-1$ patterns of $P3$. Suppose first that there are $k$ of them. This implies that there are three vertices outside of them (not necessarily consecutive). Recall that if we have a pattern~$P3$ starting from a vertex $u$, then the vertices $u-1$, $u-2$, $u-3$ and $u-4$ are all codewords. Therefore, as we have only three vertices outside of patterns~$P3$, they all have to be codewords. %If all of them are in the code, then $|K|\ge 4k+3.$ None of them can be a non-codeword, since it would belong to one of the patterns and it cannot reside to the left of any $P3$.
    Suppose then that there are $k-1$ patterns~$P3$. Now there are 10 vertices not in these patterns. If a vertex $u$ starts a pattern~$P3$ such that $u-1$ is not part of a pattern~$P3$ (indeed, such pattern has to exist), then $u-1$ is a codeword (as above) and does not belong to any pattern since none of the patterns other than $P3$ ends with four consecutive codewords. Therefore, we obtain that $7(k-1)$ vertices belongs to some pattern~$P3$, one codeword does not belong to any pattern and the rest $9$ of the vertices belong to patterns other than $P3$ (or not to any pattern). Thus, we obtain that %Again, we can assume that at least one of them is a non-codeword and belongs to a pattern which is left of some $P3$. Since $P3$ has four consecutive codewords to the left of it, there is at least one codeword (the one immediately to the left of $P3$) which is not among any patterns. But then
    $|K|\ge \frac47(7(k-1))+1+\frac35 (n-7(k-1)-1)=4k+\frac{12}{5}>4k+2$.
    %Assume there is a $SID$-code $K$ on $C_n(1,3)$ with $4k+2$ codewords. If there is no pattern $P3$ on $K$ then the density is greater than $\frac 35$ since all the sets $s_i$ defined above have all density at least $\frac35$. Since $k>1$, this means the code has at least $4k+3$ vertices, which is absurd. Hence there is at least one pattern $P3$ in the code. Assume it starts on the vertex $0$, that is, the vertices $0,1,2$ are non-codewords and the vertices $3,4,5,6$ are in the code. As the code cant be composed only of patterns $P3$ (if it is the case there is three codewords more), let us assume the vertices $\{7(k-1)+3,7(k-1)+4,\dots,7k+2\}$ are not a $P3$-pattern which means that every pattern just before the $P3$ starting at $7$ needs at most $3$ codewords at the end. For the vertices $\{0,1,2\}$ to be well identified, the vertices $\{7k-1,7k,7k+1,7k+2\}$ must be in the code, hence every pattern before $7$ ends at the latest at $7k+1$. Let $d_2$ be the density of %codewords on $\{7,8,9,\dots,7k+1\}$, we have then: %$$ d_2(7(k-1)+2)+5=4k+2 ~\Leftrightarrow~ d_2 =\frac{4k-3}{7k-5}<\frac47 $$ %and this is not possible.
    Hence, there is no self-identifying code with $4k+2$ codewords and the size of the code is at least $4k+3$. By the same argument as above, we can show that the code $K_4=\{i+7j\mid 0\leq i \leq3 , 0\leq j \leq k-1\}\cup\{7k,7k+1,7k+2\}$ works.
\item If $n=7k+5$, then we show next that the code has at least $4k+4$ codewords. It needs at least $4k+3$ codewords. %Assume there is a self-identifying code $K$ in $C_n(1,3)$ with $4k+3$ codewords.
    Let us use the sets $s_i$ of the patterns again. If there is at most $k-1$ patterns $P3$, then $|K|\ge \frac47 (7(k-1))+\frac35 (n-7(k-1))=4k+\frac{16}{5}>4k+3.$ Therefore, there must be $k$ patterns of $P3$ and five vertices outside them (not necessarily consecutive). Suppose first that these five vertices are not consecutive. Then they all must be codewords since four consecutive vertices left to any pattern~$P3$ are codewords. Suppose then that the five vertices are consecutive. This implies (with the same argument) that four of them has to be codewords. Thus, in both cases, at least four of the five vertices are codewords. Hence, we have $|K| \geq 4k+4$. %If all of the five vertices are in the code, then $|K|=4k+5.$ Hence, we can assume that at least one, say $x$, is a non-codeword. But then it belongs to one of the patterns. The pattern must be to the left of one of the patterns $P3$. However, there are at least four consecutive codewords to the left of any $P3$ pattern and thus all the five remaining vertices are consecutive, the first to the left is a non-codeword and the others are codewords. Hence, $x$ belongs to $P1a$ and $|K|>4k+3.$ Then there is no self-identifying code with $4k+3$ codewords, we need at least $4k+4$.
    As above, it is straightforward to verify that $K_6=\{i+7j\mid 0\leq i \leq3 , 0\leq j \leq k\}$ is an optimal self-identifying code with $4k+4$ vertices.
\end{itemize}
\end{proof}

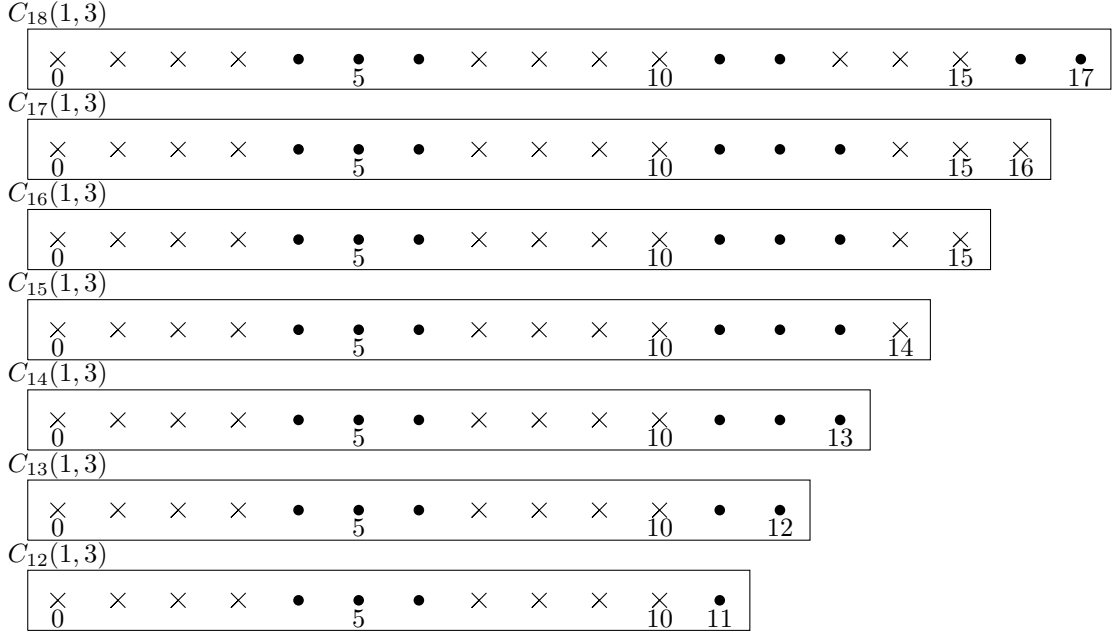
\begin{figure}%[!ht]
  \begin{center}
	\begin{tikzpicture}[scale=0.4,>=triangle 45]
%n=18
  \foreach \x in {8,10,12,22,24,32,34}
  \fill [shift={(\x,18)}] circle (5pt);
  \foreach \x in {0,2,4,6,14,16,18,20,26,28,30}
  \draw [shift={(\x,18)},color=black] (0,0)-- ++(-7.0pt,-7.0pt) -- ++(14pt,14.0pt) ++(-14.0pt,0) -- ++(14.0pt,-14.0pt);
  \foreach \x in {0,5,10,15,17}
  \draw [shift={(2*\x,18)}] node [below] {$\x$};
  \draw (0,19.5) node {$C_{18}(1,3)$};
%  \draw [shift={(0,18)}] (-1,-1) --(-1,1)--(35,1)--(35,-1)--(-1,-1);

  %n=17
  \foreach \x in {8,10,12,22,24,26}
  \fill [shift={(\x,15)}] circle (5pt);
  \foreach \x in {0,2,4,6,14,16,18,20,28,30,32}
  \draw [shift={(\x,15)},color=black] (0,0)-- ++(-7.0pt,-7.0pt) -- ++(14pt,14.0pt) ++(-14.0pt,0) -- ++(14.0pt,-14.0pt);
  \foreach \x in {0,5,10,15,16}
  \draw [shift={(2*\x,15)}] node [below] {$\x$};
  \draw (0,16.5) node {$C_{17}(1,3)$};
  
   %n=16
  \foreach \x in {8,10,12,22,24,26}
  \fill [shift={(\x,12)}] circle (5pt);
  \foreach \x in {0,2,4,6,14,16,18,20,28,30}
  \draw [shift={(\x,12)},color=black] (0,0)-- ++(-7.0pt,-7.0pt) -- ++(14pt,14.0pt) ++(-14.0pt,0) -- ++(14.0pt,-14.0pt);
  \foreach \x in {0,5,10,15}
  \draw [shift={(2*\x,12)}] node [below] {$\x$};
  \draw (0,13.5) node {$C_{16}(1,3)$};

   %n=15
  \foreach \x in {8,10,12,22,24,26}
  \fill [shift={(\x,9)}] circle (5pt);
  \foreach \x in {0,2,4,6,14,16,18,20,28}
  \draw [shift={(\x,9)},color=black] (0,0)-- ++(-7.0pt,-7.0pt) -- ++(14pt,14.0pt) ++(-14.0pt,0) -- ++(14.0pt,-14.0pt);
  \foreach \x in {0,5,10,14}
  \draw [shift={(2*\x,9)}] node [below] {$\x$};
  \draw (0,10.5) node {$C_{15}(1,3)$};

   %n=14
  \foreach \x in {8,10,12,22,24,26}
  \fill [shift={(\x,6)}] circle (5pt);
  \foreach \x in {0,2,4,6,14,16,18,20}
  \draw [shift={(\x,6)},color=black] (0,0)-- ++(-7.0pt,-7.0pt) -- ++(14pt,14.0pt) ++(-14.0pt,0) -- ++(14.0pt,-14.0pt);
  \foreach \x in {0,5,10,13}
  \draw [shift={(2*\x,6)}] node [below] {$\x$};
  \draw (0,7.5) node {$C_{14}(1,3)$};

   %n=13
  \foreach \x in {8,10,12,22,24}
  \fill [shift={(\x,3)}] circle (5pt);
  \foreach \x in {0,2,4,6,14,16,18,20}
  \draw [shift={(\x,3)},color=black] (0,0)-- ++(-7.0pt,-7.0pt) -- ++(14pt,14.0pt) ++(-14.0pt,0) -- ++(14.0pt,-14.0pt);
  \foreach \x in {0,5,10,12}
  \draw [shift={(2*\x,3)}] node [below] {$\x$};
  \draw (0,4.5) node {$C_{13}(1,3)$};

 %n=12
  \foreach \x in {8,10,12,22}
  \fill [shift={(\x,0)}] circle (5pt);
  \foreach \x in {0,2,4,6,14,16,18,20}
  \draw [shift={(\x,0)},color=black] (0,0)-- ++(-7.0pt,-7.0pt) -- ++(14pt,14.0pt) ++(-14.0pt,0) -- ++(14.0pt,-14.0pt);
  \foreach \x in {0,5,10,11}
  \draw [shift={(2*\x,0)}] node [below] {$\x$};
  \draw (0,1.5) node {$C_{12}(1,3)$};

  \foreach \x in {0,1,2,3,4,5,6}
  \draw [shift={(0,3*\x)}] (-1,-1) --(-1,1)--(2*\x+23,1)--(2*\x+23,-1)--(-1,-1);
  
\end{tikzpicture}
    \end{center}
  \caption{Optimal self-identifying codes for $C_n(1,3)$ for $n\in\{12,13,14,15,16,17,18\}$.} \label{constructionsC13}
\end{figure}

In the following theorem, we consider self-identifying codes in $C_n(1,4)$, when $n$ is odd. Recall that the cardinality of an optimal self-identifying code in $C_n(1,4)$ is $\lceil n/2 \rceil$ for even $n$ by Theorem~\ref{ThmSIDConstructions}. In particular, we show that the lower bound $\lceil n/2 \rceil$ of Corollary~\ref{sqLB} can be increased by one for odd $n$.
%\begin{theorem}%EUROCOMB
%For $k>5$ we have
%$$
%\SID(C_{2k+1}(1,4))  =  k+2.
%$$
%\end{theorem}
\begin{theorem}%EUROCOMB
If $k$ is an integer such that $k > 5$, then we have
$$
\SID(C_{2k+1}(1,4))  =  k+2.
$$
\end{theorem}
\begin{proof}
 Let $k$ and $n$ be integers such that $k > 5$ and $n = 2k+1$. Furthermore, for the lower bound, let $K$ be a self-identifying code in $C_n(1,4)$. By Corollary~\ref{sqLB}, we immediately know that $|K| \geq \lceil n/2 \rceil = k+1$. For the claim, we need to further show that $|K| = k+1$ is not possible.

Suppose first that for each $u \notin K$ we have $u-1 \in K$ and $u+1 \in K$, i.e., there does not exist consecutive non-codewords in the graph. If now $|K| = k+1$, then (without loss of generality) we can assume that the codewords are on the even vertices, i.e., $K = \{0, 2, \ldots, 2k\}$. However, this implies a contradiction since $I(K;2) = \{2,6\} \subseteq I(K;6)$. Thus, we may assume that there exist consecutive non-codewords in the graph.

Recall that we have $|I(K;c)| \geq 3$ for all $c \in K$ and $|I(K;u)| \geq 2$ for all $u \notin K$ (by Proposition~\ref{SymmetricSIDnonC}). We say that a vertex $u \in \Z_n$ is \emph{excessively covered} if $u \in K$ and $|I(K;u)| \geq 4$, or $u \notin K$ and $|I(K;u)| \geq 3$. In what follows, we first show that there exist at least three vertices that are excessively covered. Then, based on the observation, we prove that $|K| \geq k+2$. The proof now divides into the following cases depending on how many consecutive non-codewords there exist:
\begin{itemize}
\item Suppose first that there exist five or more consecutive non-codewords. If $u$ is the first one of these non-codewords, then a contradiction with Proposition~\ref{SymmetricSIDnonC}(ii) follows as $u+1 \notin K$ and $u+4 \notin K$.
\item Suppose then that there are exactly four consecutive non-codewords, say $u, u+1, u+2, u+3 \notin K$ and $u-1, u+4 \in K$. By Proposition~\ref{SymmetricSIDnonC}, we obtain that $u-4, u-3, u-2 \in K$ and $u+5, u+6, u+7 \in K$. Hence, $u$ and $u+3$ are excessively covered since they are non-codewords with at least three neighbouring codewords. Furthermore, $u+8$ is a codeword since the codeword $u+4$ has to be covered by at least three codewords. Now, if $u+9 \in K$, then the codeword $u+5$ is excessively covered since $|I(K;u+5)| \geq 4$. On the other hand, if $u+9 \notin K$, then $u+9$ is excessively covered since $u+5$ and $u+8$ belong to $K$ as well as at least one of the vertices $u+10$ and $u+13$.
\item Suppose that there are exactly three consecutive non-codewords, say $u, u+1, u+2 \notin K$ and $u-1, u+3 \in K$. As above, we deduce that $u-4, u-3, u-2 \in K$ and $u+4, u+5, u+6 \in K$. Similar to the previous case, we immediately obtain that $u$ and $u+2$ are excessively covered. If $u+7 \in K$, then $u+3$ is excessively covered since $|I(K;u+3)| \geq 4$. On the other hand, if $u+7 \notin K$, then $u+7$ is excessively covered (as the vertex $u+9$ in the previous case). Thus, we have three excessively covered vertices.
\item Suppose that there are exactly two consecutive non-codewords, say $u, u+1 \notin K$ and $u-1, u+2 \in K$. As above, we first obtain that $u-4, u-3 \in K$ and $u+4, u+5 \in K$. Using similar arguments as earlier, we immediately obtain that $u$ and $u+1$ are excessively covered. Furthermore, since the codewords $u-1$, $u+2$ and $u+4$ all belong to  $I(K;u+3)$, the vertex $u+3$ is excessively covered regardless whether it is a codeword or a non-codeword. Thus, we have three excessively covered vertices.
%\item Suppose that there are exactly two consecutive non-codewords, say $u, u+1 \notin K$ and $u-1, u+2 \in K$. As above, we first obtain that $u-4, u-3 \in K$ and $u+4, u+5 \in K$. Using similar arguments as earlier, we immediately obtain that $u$ and $u+1$ are excessively covered. If $u+3 \in K$, then $u+3$ is excessively covered (as $\{u-1, u+2, u+3, u+4\} \subseteq I(K;u+3)$) and we are done. Analogously, we are done if $u-2 \in K$. Hence, we may assume that $u-2$ and $u+3$ do not belong to $K$. However, this implies a contradiction with Proposition~\ref{SymmetricSIDnonC} for the vertex $u+2$; indeed, neither of $u-2$ and $u+1$ belong to $K$. Thus, we have three excessively covered vertices.
\end{itemize}

As stated earlier, we have $|I(K;c)| \geq 3$ for all $c \in K$ and $|I(K;u)| \geq 2$ for all $u \notin K$. In addition, we have shown that at least three vertices are excessively covered, i.e., covered more than what is required here. Therefore, by counting in two ways the pairs $c \in K$ and $u \in \Z_n$ such that $u \in N[c]$, we obtain the following inequality:
\[
5|K| \geq 3|K| + 2(n-|K|) +3 ~\Leftrightarrow~ |K| \geq \left\lceil \frac{2n+3}{4} \right\rceil = k+2 \text{.}
\]
Thus, in conclusion, we have shown that $|K| \geq k+2$.

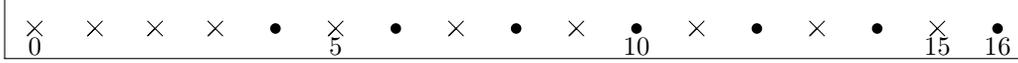
\begin{figure}%[!ht]
\begin{center}
\begin{tikzpicture}[scale=0.4,line cap=round,line join=round,>=triangle 45,x=1.0cm,y=1.0cm]
%\clip(-1.3,-1.1) rectangle (33.5,1.1);
  \foreach \x in {8,12,16,20,24,28,32}
  \fill [shift={(\x,0)}] circle (5pt);
  \foreach \x in {0,2,4,6,10,14,18,22,26,30}
  \draw [shift={(\x,0)},color=black] (0,0)-- ++(-7.0pt,-7.0pt) -- ++(14pt,14.0pt) ++(-14.0pt,0) -- ++(14.0pt,-14.0pt);
  \foreach \x in {0,5,10,15,16}
  \draw [shift={(2*\x,0)}] node [below] {$\x$};
%  \draw (0,2) node {$C_{17}(1,7)$};
  \draw (-1,-1) --(-1,1)--(33,1)--(33,-1)--(-1,-1);

\end{tikzpicture}
\end{center}
\caption{Optimal self-identifying code on $C_{17}(1,4)$.}\label{constructionC14}
\end{figure}
For the construction attaining the lower bound, we denote $K_1 = \{0,2\} \cup \{i \in \Z_n \mid i \text{ is odd} \}$. The code $K_1$ is illustrated in Figure~\ref{constructionC14} (when $n=17$). Clearly, $K_1$ contains $k+2$ codewords. Furthermore, it is self-identifying in $C_{n}(1,4)$. Indeed, for $v\in \{4,\dots,n-1\}$, we have $I(v) = \{v-4,v,v+4\}$ if $v \in K_1$, and $I(v)$ contains $\{v-1,v+1\}$ if $v \notin K_1$. It is also straightforward to verify that the codewords in $I(v)$ intersect uniquely in $v$ for $v = 0,1,2,3$. Hence, $K_1$ is an optimal self-identifying code.
\end{proof}

In the following theorem, we give optimal self-identifying codes for $C_n(1,n/2)$ for $n$ even.
 Let $G=(V,E)$ be a graph and $K\subseteq V.$ The \emph{minimum distance} of a code $K$ is defined via $$d_{\textrm{min}}(K)=\min_{x,y\in K, x\neq y}d_G(x,y).$$
 We call $K\subseteq V$ a $1$-\emph{error correcting} code, if $d_{\textrm{min}}(K)\ge 3$. If $K$ is $1$-error correcting, then $I(G,K;x)=\{x\}$ for all $x\in K$ and $N(G;x)\cap N(G;y)=\emptyset$ for all distinct $x,y\in K.$ Moreover, if $G$ is $r$-regular, then a $1$-error correcting code must satisfy the \emph{sphere packing bound}: $|K|\le |V|/(r+1).$

\begin{theorem} %RUFIDIM
  Let $k\geq 5$. The optimal cardinality of self-identifying code in $C_{2k}(1,k)$ is as follows:
  $$\displaystyle
  \SID(C_{2k}(1,k)) =\left\{
  \begin{array}{ll}
    \lceil\frac{4k}{3}\rceil & \text{ if } k\equiv 0,1 \pmod{3} \\

    \lceil\frac{4k}{3}\rceil+1 & \text{ if } k\equiv 2\pmod{3}
  \end{array}
  \right.
  $$
\end{theorem}
\begin{proof}
  Let $k\geq 5$ and $n=2k$. We study self-identifying codes in the graph $C_n(1,k)$. For all $x\in V=\{0,\dots,n-1\}$ the closed neighbourhood of $x$ is $N[x]=\{x-k,x-1,x,x+1,x+k\} = \{x-k,x-1,x,x+1\}$ as $x-k\equiv x+k \bmod n$. Let $K$ be a self-identifying code. Now it is easy to see that the $I$-set $I(x)$ contains $\{x-1,x+1\}$ for all $x \notin K$. %It is easy to check that if $K$ is a self-identifying code, then if $x$ is a non-codeword, the set  $I(x)$ contains $\{x-1,x+1\}$.
  Therefore, non-codewords are always surrounded by codewords (in the cycle $C_n(1)$). Furthermore,  a codeword $v\in K$ cannot be surrounded by two non-codewords (in $C_n(1)$). Indeed it would imply that $I(v)=\{v,v+k\}\subseteq I(v+k)$. Hence the minimum distance $d_{\textrm{min}}(V\setminus K)\ge 3$ and the set of non-codewords forms a $1$-error correcting code in the cycle $C_n(1)$. Consequently, by the sphere packing bound $|V\setminus K|\le  n/3.$ This observation yields $|K|\ge \lceil\frac{2}{3}n\rceil$, which gives the claimed lower bound for $k\equiv 0,1 \pmod{3}.$ The constructions attaining the bound are given next:
  \begin{itemize}
  \item Let $k\equiv 0 \pmod{3}$. The code $K_1=\{v\in \{0,\dots, n-1\}\mid v\not\equiv 2\pmod{3}\}$ is self-identifying. Indeed, if $x\equiv 1\pmod{3}$, then $\{x-1,x+1\}\subseteq I(x)$ and their intersection $N[x-1]\cap N[x+1]$ contains only $x$. If $x\equiv 0,2\pmod{3}$, then $\{x,x-k\}\subseteq I(x)$ and $I(x)$ also contains a third codeword and the intersection of them equals $x$.

  \item Let $k\equiv 1 \pmod{3}$. If $S_2=\{v\in \{0,\dots, k-1\}\mid v\not\equiv 2\pmod{3} \}$, then the code $K_2=S_2\cup \{s+k\mid s\in S_2\}$ is self-identifying.
      %the code $K_2=\{v\in \{0,\dots, n-1\}\mid v \equiv s\pmod{k} \text{ for some } s\in S\}$ is self-identifying.
      Indeed, every non-codeword $v$ we have $\{v-1,v+1\}\subseteq I(v)$ and we are done. For every codeword  $v$ the $I(v)$ contains either $\{v,v- 1,v+k\}$ or $\{v,v+ 1,v+k\}$ and we are done (see Figure \ref{tightness2k}).
     \end{itemize}
  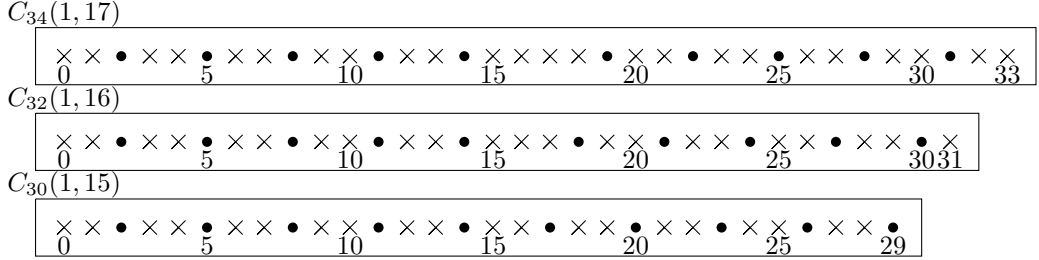
\begin{figure}[!ht]
      \begin{center}
    \begin{tikzpicture}[scale=0.38,>=triangle 45]
\clip(-4,-1.1) rectangle (35.5,7.9);
  \foreach \x in {2,5,8,11,14}
           {\fill [shift={(\x,6)}] circle (5pt);
            \fill [shift={(17+\x,6)}] circle (5pt);
           } 
  \foreach \x in {0,1,3,4,6,7,9,10,12,13,15,16}
           {\draw [shift={(\x,6)},color=black] (0,0)-- ++(-7.0pt,-7.0pt) -- ++(14pt,14.0pt) ++(-14.0pt,0) -- ++(14.0pt,-14.0pt);
            \draw [shift={(17+\x,6)},color=black] (0,0)-- ++(-7.0pt,-7.0pt) -- ++(14pt,14.0pt) ++(-14.0pt,0) -- ++(14.0pt,-14.0pt);
           }
  \foreach \x in {0,5,10,15,20,25,30,33}
    \draw [shift={(\x,6)}] node [below] {$\x$};
  \draw (0,7.5) node {$C_{34}(1,17)$};
  \draw (-1,7) --(-1,5)--(34,5)--(34,7)--(-1,7);

  \foreach \x in {2,5,8,11,14}
           {\fill [shift={(\x,3)}] circle (5pt);
            \fill [shift={(16+\x,3)}] circle (5pt);
           }
  \foreach \x in {0,1,3,4,6,7,9,10,12,13,15}
           {\draw [shift={(\x,3)},color=black] (0,0)-- ++(-7.0pt,-7.0pt) -- ++(14pt,14.0pt) ++(-14.0pt,0) -- ++(14.0pt,-14.0pt);
            \draw [shift={(16+\x,3)},color=black] (0,0)-- ++(-7.0pt,-7.0pt) -- ++(14pt,14.0pt) ++(-14.0pt,0) -- ++(14.0pt,-14.0pt);
           }
  \foreach \x in {0,5,10,15,20,25,30,31}
    \draw [shift={(\x,3)}] node [below] {$\x$};
  \draw (0,4.5) node {$C_{32}(1,16)$};
  \draw (-1,4) --(-1,2)--(32,2)--(32,4)--(-1,4);

  \foreach \x in {2,5,8,11,14}
           {\fill [shift={(\x,0)}] circle (5pt);
            \fill [shift={(15+\x,0)}] circle (5pt);
           }
  \foreach \x in {0,1,3,4,6,7,9,10,12,13}
           {\draw [shift={(\x,0)},color=black] (0,0)-- ++(-7.0pt,-7.0pt) -- ++(14pt,14.0pt) ++(-14.0pt,0) -- ++(14.0pt,-14.0pt);
            \draw [shift={(15+\x,0)},color=black] (0,0)-- ++(-7.0pt,-7.0pt) -- ++(14pt,14.0pt) ++(-14.0pt,0) -- ++(14.0pt,-14.0pt);
           }
  \foreach \x in {0,5,10,15,20,25,29}
    \draw [shift={(\x,0)}] node [below] {$\x$};
  \draw (0,1.5) node {$C_{30}(1,15)$};
  \draw (-1,1) --(-1,-1)--(30,-1)--(30,1)--(-1,1);

\end{tikzpicture}
    \end{center}
     \caption{Examples of optimal self-identifying codes for $k=15,16,17$.}\label{tightness2k}
    \end{figure}
   It suffices to consider the case  $k\equiv 2 \pmod{3}.$ We show that the cardinality of $K$ must be greater than $\lceil \frac{2}{3}n \rceil= (2n+1)/3.$ Suppose to the contrary that $|K|= (2n+1)/3.$ Then there are $(n-1)/3$ non-codewords. Since $V\setminus K$ is $1$-error correcting of cardinality $(n-1)/3$, there is exactly one vertex $y\in V$ outside the disjoint sets $N[C_n(1);v]$ for $v\in V\setminus K$. In other words, once $y$ (clearly, $y\in K$) is given, then we know the code $K$ without ambiguity.  Without loss of generality, let $y=0$ and thus  the code is $K=\{v\in \Z_n\mid v\not\equiv 2 \pmod{3}\}.$
   However, with this code we have $N[C_n(1,k);3]=\{2,3,4,3+k\}$ but $2$ and $3+k$ are not in the code. Thus we have $\{3,4\} = I(C_n(1,k);3)\subseteq I(C_n(1,k);4) = \{3,4,4+k\}$ and $K$ cannot be self-identifying. We conclude that every self-identifying code in $C_n(1,k)$ needs at least $\lceil \frac{2}{3}n \rceil+1$ codewords.

    Denote $S_3=\{v\in \{0,\dots, k-1\}\mid v\not\equiv 2\pmod{3} \}$. The construction attaining the bound $\lceil \frac{2}{3}n \rceil+1$ is $K_3=S_3\cup \{s+k\mid s\in S_3\}$. The code $K_3$ is self-identifying. Indeed, for every non-codeword $v$ we have $\{v-1,v+1\}\subseteq I(v)$ and for every codeword  $v$ either $\{v,v- 1,v+k\}\subseteq I(v)$ or $\{v,v+ 1,v+k\}\subseteq I(v)$ and we are done.
\end{proof}

\section{On attaining some lower bounds}

Let us first introduce two basic result on identifying and self-identifying codes, which we need in Theorem~\ref{nonreachableThm}. The first bound considering identifying codes is well-known (see \cite{kcl}), but we add the proof for completeness.

\begin{theorem} \label{ThmOptimalChar}
Let $k$ be an integer such that $k \geq 2$ and $G = (V,E)$ be a finite $k$-regular graph.
\begin{itemize}
\item[(i)] We have the following lower bound for the cardinality of an optimal identifying code:
    \[
    \ID(G) \geq \left\lceil \frac{2|V|}{k+2} \right\rceil \text{.}
    \]
    Moreover, there exists an identifying code $C$ in $G$ such that $|C| = 2|V|/(k+2)$ if and only if there exist exactly $|C|$ vertices $u \in V$ such that $|I(C;u)| = 1$ and for all other vertices $v \in V$ we have $|I(C;v)| = 2$.
%\item[(ii)] We have the following lower bound for the cardinality of an optimal locating-dominating code:
%    \[
%    \LD(G) \geq \left\lceil \frac{2|V|}{k+3} \right\rceil \text{.}
%    \]
%    Moreover, there exists a locating-dominating code $C$ in $G$ such that $|C| = 2|V|/(k+3)$ if and only if there exist exactly $2|C|$ vertices $u \in V$ such that $|I(C;u)| = 1$ and for all other vertices $v \in V$ we have $|I(C;v)| = 2$.
\item[(ii)] We have the following lower bound for the cardinality of an optimal self-identifying code:
    \[
    \SID(G) \geq \left\lceil \frac{2|V|}{k} \right\rceil \text{.}
    \]
    Moreover, there exists a self-identifying code $C$ in $G$ such that $|C| = 2|V|/k$ if and only if $|I(C;u)| = 3$ for all $u \in C$ and $|I(C;v)| = 2$ for all $v \in V \setminus C$.
\end{itemize}
\end{theorem}
\begin{proof}
(i) Let first $C$ be an identifying code in $G$. Observe then that there are at most $|C|$ identifying sets with exactly one codeword since the code $C$ is identifying. Hence, all the other $|V| - |C|$ identifying sets have at least two codewords. Therefore, by counting in two ways the pairs $c \in C$ and $u \in V$ such that $u \in N[c]$, we obtain the following inequality:
\[
|C|(k+1)\geq |C|+2(|V|-|C|) ~\Leftrightarrow~ |C|\geq \frac{2|V|}{k+2} \text{.}
\]
Moreover, by the previous observations, $|C| = 2|V|/(k+2)$ if and only if there exist exactly $|C|$ vertices $u \in V$ such that $|I(C;u)| = 1$ and for all other vertices $v \in V$ we have $|I(C;v)| = 2$.

%(ii) Let then $C$ be a locating-dominating code in $G$. Observe that there are at most $2|C|$ identifying sets with exactly one codeword since the code $C$ is locating-dominating. Hence, all the other $|V| - 2|C|$ identifying sets have at least two codewords. Therefore, by a simple double counting argument, we obtain the following inequality:
%\[
%|C|(k+1)\geq 2|C|+2(|V|-2|C|) ~\Leftrightarrow~ |C|\geq \frac{2|V|}{k+3} \text{.}
%\]
%Moreover, by the previous observations, $|C| = 2|V|/(k+3)$ if and only if there exist exactly $2|C|$ vertices $u \in V$ such that $|I(C;u)| = 1$ and for all other vertices $v \in V$ we have $|I(C;v)| = 2$.

(ii) Finally, let $C$ be a self-identifying code in $G$. Observe then that $N(u) \cap C$ has at least two codewords of $C$ for all $u \in V$ since the code $C$ is self-identifying. Hence, for each $u \in C$ we have $|I(C;u)| \geq 3$ and for each $u \in V \setminus C$ we have $|I(C;u)| \geq 2$. Therefore, by a similar double counting argument as above, we obtain the following inequality:
\[
|C|(k+1)\geq 3|C|+2(|V|-|C|) ~\Leftrightarrow~ |C|\geq \frac{2|V|}{k} \text{.}
\]
Moreover, by the previous observations, $|C| = 2|V|/k$ if and only if $|I(C;u)| = 3$ for all $u \in C$ and $|I(C;v)| = 2$ for all $v \in V \setminus C$.
\end{proof}

The previous theorem gives lower bounds for circulant graphs as they are also regular. In the following theorem, we show that the exact bounds (above) cannot be attained in circulant graphs for identifying and self-identifying codes.
\begin{theorem} \label{nonreachableThm}
Let $n$, $r$ and $d_2, \ldots, d_r$ be integers such that $r \geq 3$ and $1<d_2<\dots<d_r\le n/2$.
\begin{itemize}
\item[(i)] Then there does not exist any identifying code $C$ in $C_n(1,d_2,\dots,d_r)$ such that $|C| = n/(r+1)$.
\item[(ii)] Then there does not exist any self-identifying code $C$ in $C_n(1,d_2,\dots,d_r)$ such that $|C| = n/r$.
\end{itemize}
\end{theorem}
\begin{proof}
(i) Let $C$ be an identifying code in $C_n(1,d_2,\dots,d_r)$ such that $|C| = n/(r+1)$. By Theorem~\ref{ThmOptimalChar}, it is possible if and only if there are exactly $|C|$ vertices $x_1,\dots,x_{|C|}$ such that $|I(x_i)|=1$ and the rest of the vertices have identifying sets with exactly two vertices. If there exists a vertex of $C$, say $u$, such that $|I(C;u)|=2$, then we have $I(C;u)=\{u,v\}$ and $I(C;u)=I(C;v)$ as all identifying sets have at most two codewords (a contradiction). Hence, the codewords of $C$ are the vertices $x_1,\dots,x_{|C|}$. Therefore, in particular, we have $I(C;x_1)=\{x_1\}$ implying that $x_1+1\notin C$ and $|I(C;x_1+1)|=2$. If $I(C;x_1+1)=\{x_1,x_1+1\pm d_i\}$, then the vertex $v=x_1\pm d_i$ contains $\{x_1, x_1+1\pm d_i\}$ in its identifying set. Therefore, a contradiction follows as above. Hence, it has to be that $I(C;x_1+1)=\{x_1,x_1+2\}$. Now, because $x_1 + 2 \in C$, we have $I(C;x_1+2) = \{x_1+2\}$. Then, using similar arguments as above, we obtain that $x_1+3 \notin C$ and $I(C;x_1+3)=\{x_1+2,x_1+4\}$. Thus, by continuing this process, we obtain that every other vertex of $C_n(1,d_2,\dots,d_r)$ is a codeword. Clearly, this leads to a contradiction with the chosen cardinality of $C$. Thus, we conclude that $\ID(C_n(1,d_2,\dots,d_r)) > n/(r+1) $.

(ii) Let then $C$ be a self-identifying code in $C_n(1,d_2,\dots,d_r)$ such that $|C| = n/r$. By Theorem~\ref{ThmOptimalChar}, it is possible if and only if for each $u \in \Z_n$ we have $|N(u) \cap C| = 2$, i.e., $u$ has exactly two codewords of $C$ in its open neighbourhood. Clearly, there exists a vertex $x \in \Z_n \setminus C$ such that $x-1 \in C$. Using similar arguments as in the case of identifying codes, we obtain that $I(C;x) = \{x-1, x+1\}$. We continue to the right and use the same argument for each non-codeword that comes along. Hence, for any non-codeword $y \in \Z_n \setminus C$, we have $I(C;y) = \{y-1, y+1\}$. This further implies that $|C| \geq n/2$ which is a contradiction with the chosen cardinality of $C$, since $r\geq 3$. Thus, we conclude that $\SID(C_n(1,d_2,\dots,d_r)) > n/r$.
\end{proof}

In the case of identifying codes, we immediately obtain the following corollary.
\begin{corollary} \label{CorollaryUnreachable}
If $n$ and $d$ are positive integers such that $d \geq 4$ and $d \le n/2$, then we have
\[
\ID(C_n(1,d-1,d)) > \frac{n}{4} \text{.}
\]
\end{corollary}
Thus, the bound announced in Corollary \ref{triLB} is never reached, but it is best possible since there is a sequence of codes on the circulant graphs $C_n(1,d-1,d)$ tending to this bound as proved in Theorem \ref{Triangular_tightness}.\\

\section{Appendix}
%\subsection{Proof of the Theoreom~\ref{KingThm}}
\emph{The proof of the Theorem~\ref{KingThm}(ii):}

Let $d\ge 15$, $d\equiv 3 \pmod{6}$ and $n=3d-9$. Notice that $n\equiv 0 \pmod{6}.$ We divide the vertices of the circulant graph into three sections denoted by $A_1=\{0,1,2\dots, d-1\}$, $A_2=\{d,d+1,\dots, 2d-1\}$ and $A_3=\{0,1,\dots,n-1\}\setminus(A_1\cup A_2).$  We will first consider the code
$$
C_d=\{v\mid v\in (A_1\cup A_3), v\equiv 5 \pmod{6}\} \cup \{v\mid v\in A_2, v\equiv 0,4 \pmod{6}\}.
$$
%$$\cup\{v\mid v\in A_3,v\equiv 5 \pmod{5}\}$$%\cup\{0,d,2d+1\}$$
Using this code we can construct (by adding later two more codewords) an identifying code in $C_{n}(1,d-1,d,d+1)$.
The ratio $|C_d|/n$ tends to $2/9$ as $d$ tends to infinity. First we exclude some `borderline' vertices from the three sections and denote $A_1'=A_1\setminus\{0,1,2,3,4,5,6,7,8,9,d-1\}$, $A_2'=A_2\setminus\{d,2d-1\}$ and $A_3'=A_3\setminus\{2d\}.$ We consider the borderline vertices later.
It is straightforward to check that the $I$-sets with regard to the code $C_d$ are as follows for $x\in A_1'\cup A_2'\cup A_3'$:

$$
\begin{array}{cccc}
  x\in A_1' & I(x) & d(c_1,c_2) & I(x)\textrm{ mod }6 \\ \hline
  \equiv 0 \textrm{ mod } 6 & \{x-1,x+d+1\} & d+2 & 4,5 \\
  1 & \{x-d+1,x+d\} & d-8 & 4,5\\
  2 & \{x-d,x+d-1,x+d+1\} &  & 0,4,5 \\
  3 & \{x-d-1,x+d\} & d-10 & 0,5\\
  4 & \{x+1,x+d-1\}  & d-2 & 0,5\\
  5 & \{x\} & &
\end{array}
$$

$$
\begin{array}{cccc}
  x\in A_2' & I(x) & d(c_1,c_2) & I(x)\textrm{ mod }6 \\ \hline
  \equiv 0 \textrm{ mod }6 & \{x\} &  &  \\
  1 & \{x-d+1,x-1,x+d+1\} & & 0,5,5\\
  2 & \{x-d,x+d\} & 2d & \\
  3 & \{x-d-1,x+1,x+d-1\} &  & 4,5,5\\
  4 & \{x\}  &  & \\
  5 & \{x-1,x+1\} & 2 &
\end{array}
$$

$$
\begin{array}{cccc}
  x\in A_3' & I(x) & d(c_1,c_2) & I(x) \textrm{ mod } 6 \\ \hline
  \equiv 0\textrm{ mod }6 & \{x-d+1,x-1\} & d-2 & 4,5 \\
  1 & \{x-d,x+d+1\} & d-10 & 4,5\\
  2 & \{x-d-1,x-d+1,x+d\} &  & 0,4,5  \\
  3 & \{x-d,x+d-1\} & d-8 & 0,5\\
  4 & \{x-d-1,x+1\}  & d+2 & 0,5\\
  5 & \{x\} & &
\end{array}
$$

Let us compare these $I$-sets (that is, when $x\in A_1'\cup A_2'\cup A_3'$).
Clearly, the $I$-sets of size one are distinguished. Consider then the $I$-sets of size two. In the tables above, one can found the distances $c_1-c_2$ of  the codewords in $I(x)$ with $c_1>c_2$. If the distance is different, the $I$-sets cannot be the same.  For those, which have the same distance, the $c_1 \pmod{6}$ and $c_2 \pmod{6}$ are different as shown in the table, and the $I$-sets again cannot be the same. Let us study the $I$-sets of size three then. According to the tables, the codewords in the $I$-sets are different modulo 6 unless $x\in A_1'$ where $x\equiv 2 \pmod 6$ and $y\in A_3'$ where  $y\equiv 2\pmod{6}$. However, now $I(y)$ has distance 2 between its two largest codewords, but  $I(x)$ has corresponding distance $d-10.$ Consequently, $I(x)\neq I(y).$

For the rest of the vertices (i.e., the borderline vertices $x\notin A_1'\cup A_2'\cup A_3'$) we get the following $I$-sets: $I(0)=\{d+1,2d-8,3d-10\}$, $I(1)=\{d+1,2d-8\}$, $I(2)=\{d+1,d+3,2d-8,2d-6\}$, $I(3)=\{d+3,2d-6\}$, $I(4)=\{5,d+3,2d-6\}$, $I(5)=\{5\}$, $I(6)=\{5,d+7,2d-2\}$, $I(7)=\{d+7,2d-2\}$, $I(8)=\{d+7,d+9,2d-2\}$, $I(9)=\{d+9\}$, $I(d-1)=\{2d-2,3d-10\},$ $I(d)=\{d+1,3d-10\}$, $I(2d-1)=\{2d-2\}$ and $I(2d)=\{d+1\}$. It is straightforward to check (considering sizes of $I$-sets, codewords modulo 6 in $I$-sets and their distances) that we have exactly the following non-distinguished $I$-sets: $I(9)=I(d+9)$, $I(d-1)=I(d-2)$, $I(d+1)=I(2d)$ and $I(2d-2)=I(2d-1)$. We add two  more codewords, namely, $0$ and $2d$ to the code $C_d$ to avoid these same $I$-sets. Denote $C_d'=C_d\cup \{0,2d\}.$ We should bear in mind that if $I(C_d;x)\neq I(C_d;y)$, then also $I(C_d';x)\neq I(C_d';y).$ Now we have (with respect to $C_d'$) that $2d\in I(9)\setminus I(d+9)$, $0\in I(d-1)\setminus I(d-2)$, $2d\in I(2d-1)\setminus I(2d-2)$ and $0\in I(d+1)\setminus I(2d)$. Therefore, $C_d'$ is an identifying code and the proof is completed.
%\subsection{Figure of Theorem~\ref{tightness_1d}$(i)$}

%\bibliographystyle{abbrv}
%\bibliography{../../Identifying/Bibliography/refbib}
%\bibliography{refbibVille2017}

\end{document}